\documentclass{aa}  

\usepackage{graphicx}
\usepackage[varg]{txfonts}
\usepackage{pdflscape}
\begin{document}
   \title{A search for diffuse bands in fullerene planetary
nebulae: evidence of diffuse circumstellar bands}

   \titlerunning{Diffuse bands in fullerene PNe}

   \author{J. J.  D\'{\i}az-Luis\inst{1,2}, D. A.
   Garc\'{\i}a-Hern\'andez\inst{1,2}, N. Kameswara Rao\inst{1,2,3}, A. Manchado\inst{1,2,4}, \and F. Cataldo\inst{5,6} }
	  
\authorrunning{D\'{\i}az-Luis et al.}

   \institute{Instituto de Astrof\'{\i}sica de Canarias, C/ Via L\'actea s/n, E$-$38205 La Laguna, Spain \email{jdiaz@iac.es, agarcia@iac.es}
         \and Departamento de Astrof\'{\i}sica, Universidad de La Laguna (ULL), E$-$38206 La Laguna, Spain
	 \and Indian Institute of Astrophysics, Bangalore 560034, India; nkrao@iiap.res.in
         \and Consejo Superior de Investigaciones Cient\'{\i}ficas, Madrid, Spain
         \and INAF- Osservatorio Astrofisico di Catania, Via S. Sofia 78, Catania 95123, Italy
         \and Actinium Chemical Research srl, Via Casilina 1626/A, 00133 Rome, Italy
             }

   \date{Received xx, 2014; accepted xx x, 2014}

 
\abstract
{Large fullerenes and fullerene-based molecules have been proposed as carriers
of diffuse interstellar bands (DIBs). The recent detection of the most common
fullerenes (C$_{60}$ and C$_{70}$) around some planetary nebulae (PNe) now
enable us to study the DIBs towards fullerene-rich space environments. We search
DIBs in the optical spectra towards three fullerene-containing PNe (Tc 1, M
1-20, and IC 418). Special attention is given to DIBs which are found to be
unusually intense towards these fullerene sources. In particular, an unusually
strong 4428 \AA~absorption feature is a common charateristic of fullerene PNe.
Similar to Tc 1, the strongest optical bands of neutral C$_{60}$ are not
detected towards IC 418. Our high-quality (S/N $>$ 300) spectra for PN Tc 1,
together with its large radial velocity, permit us to search for the presence of
diffuse bands of circumstellar origin, which we refer to as diffuse circumstellar
bands (DCBs). We report the first tentative detection of two DCBs at 4428 and
5780 \AA~in the fullerene-rich circumstellar environment around the PN Tc 1.
Laboratory and theoretical studies of fullerenes in their multifarious
manifestations (carbon onions, fullerene clusters, or even complex species
formed by fullerenes and other molecules like PAHs or metals) may help solve
the mystery of some of the diffuse band carriers.}


\maketitle


\section{Introduction}

Identification of the carriers of the diffuse interstellar bands (DIBs) has
been very elusive since they were first discovered by Heger (1922), who
first noted their stationary nature as observed towards a spectroscopic binary,
indicating that their origin was not stellar but rather interstellar. Since then,
more than 380 bands have been identified (e.g., Hobbs et al. 2008), and they have
been associated to the interstellar medium (ISM) because their strengths show a
positive relationship with the observed extinction (Merrill \& Wilson 1936), as
well as to the neutral sodium column density (Herbig 1993). Most of the DIBs are
located in the 4000 to 10000 \AA\ wavelength range\footnote{Geballe et al.
(2011) report 13 newly discovered DIBs in the near-infrared region at the
H-band (1.5-1.8 micrometre interval) on high-extinction sightlines towards stars
in the Galactic centre.}. Different complex carbon-based molecules - e.g.,
carbon chains, polycyclic aromatic hydrocarbons (PAHs), and fullerenes - have
been proposed as DIB's carriers (see, e.g., Cox 2011 for a review). 

The link between PAHs and DIBs was made by Crawford et al. (1985), Leger \&
d'Hendecourt (1985), and Van der Zwet \& Allamandola (1985). Apart from having
(electronic) transitions in the UV, optical, and near infrared, PAHs are also
very resistant to UV radiation. Thus, the expected high abundance of PAHs in
space, their optical absorption spectrum, and the presence of substructure in
the DIB profiles, seem to give support for the PAH-DIB hypothesis (see, e.g.,
Salama et al. 1999, Cox 2011). However, a potential problem with the PAH-DIB
hypothesis is the lack of interstellar bands in the UV part of the astronomical
spectra (Snow \& McCall 2006; Snow \& Destree 2011), where strong PAH
transitions are expected (see, e.g., Tielens 2008).

Fullerenes and fullerene-related molecules (Kroto et al. 1985) are presented as
an alternative to the PAH-DIB hypothesis. The remarkable stability of fullerenes
against intense radiation (e.g., Cataldo et al. 2009) suggests that fullerenes
may be present in the ISM. The most common fullerenes (C$_{60}$ and C$_{70}$)
have recently been detected in a variety of space environments, such as planetary
nebulae (PNe) (Cami et al. 2010; Garc\'{\i}a-Hern\'andez et al. 2010, 2011a,
2012a), reflection nebulae (Sellgren et al. 2010), a proto-PN (Zhang \& Kwok
2011), and the two least H-deficient R Coronae Borealis stars
(Garc\'{\i}a-Hern\'andez et al. 2011b,c). The recent detection of fullerenes in
PNe with normal H-abundances (Garc\'{\i}a-Hern\'andez et al. 2010) indicates
that fullerenes are common around evolved stars and that they should be
widespread in the ISM. Indeed, the 9577 and 9632 \AA\ DIBs observed in a few
reddened stars lie near two electronic transitions of the C$_{60}$$^+$ cation
observed in rare gas matrices (Foing \& Ehrenfreund 1994). More recently,
Iglesias-Groth \& Esposito (2013) have reported the detection of the 9577 and 9632 \AA\
DIBs in the fullerene-containing proto-PN IRAS 01005$+$7910, and they suggest the
C$_{60}$$^+$ cation as their carrier. However, a confirmation of the proposal
that C$_{60}$$^+$ could be the carrier of these two DIBs still awaits
spectroscopic gas-phase C$_{60}$$^+$ laboratory data. The detection of
C$_{60}$$^+$ through its infrared vibrational bands in the NGC 7023 reflection
nebula with the {\it Spitzer space telescope} could support the idea of
C$_{60}$$^+$ being a DIB carrier (Bern\'{e} et al. 2013, 2014), but these
C$_{60}$$^+$ infrared bands are not seen in the {\it Spitzer} spectrum
of the proto-PN IRAS 01005$+$7910 (see, e.g., Zhang \& Kwok 2011).

At present, very little is known about the presence of the DIB carriers in other
astrophysical environments (e.g., Cox 2011). If the DIBs arise from large gas
phase molecules, such as PAHs and fullerenes, then they are also expected to be
present in other carbon-rich space environments like circumstellar shells around
stars. Diffuse circumstellar bands (DCBs) in absorption have been unsuccessfully
studied for more than 40 years (Seab 1995)\footnote{We note that some diffuse
bands in emission have been previously seen in a proto-PN (the Red Rectangle)
and in the R Coronae Borealis star V854 Cen (see, e.g., Scarrott et al. 1992; Rao
\& Lambert 1993).}. DCBs are absent in the dusty circumstellar envelopes (with or
without PAH-like features) of AGB/post-AGB stars, as well as in the atmospheres
of cool stars and Herbig Ae/Be stars (Seab 1995; Cox 2011; Luna et al. 2008).
Thus, the conventional wisdom is that there are no diffuse bands in
circumstellar environments. The unambiguous detection of DCBs would have a
strong impact on diffuse bands theories; for example, they can be compared to the
presence of the proposed diffuse band carriers mentioned above. The main
difficulty to detect DCBs is to distinguish them from the DIBs (Seab 1995; Cox
2011). This distinction can only be made by measuring the radial velocities of
the circumstellar and interstellar components. Here we search for the possible
presence of DCBs in a selected sample of three fullerene-containing PNe. 

In Garc\'{\i}a-Hern\'andez \& D\'{\i}az-Luis (2013), we presented some of the
new results for DIBs towards the fullerene PNe Tc 1 and M 1-20, and the
very broad 4428 \AA\ DIB was found to be unusually intense (based on the
measured equivalent widths) towards both Tc 1 and M 1-20. The speculation was
offered that the unusually strong 4428 \AA\ DIB towards fullerene PNe may
be related to the presence of larger fullerenes and buckyonions in their
circumstellar envelopes. However, Garc\'{\i}a-Hern\'andez \&
D\'{\i}az-Luis (2013) did not carry out any radial velocity analysis, something
that is mandatory for confirming a circumstellar origin. 

In this paper, we present a detailed DIB radial velocity analysis and a complete search
of diffuse bands towards three PNe (Tc 1, M 1-20, and IC 418) containing
fullerenes and accompanied (or not) by PAH molecules. A summary of the optical
spectroscopic observations is presented in Section 2. Section 3 gives a
complete analysis of the DIBs towards fullerene PNe, including the normal
DIBs most commonly found in the ISM and a few unusually strong DIBs. Section 4
presents our search for DCBs in fullerene PNe and their detection in PN Tc 1.
Sections 5 and 6 discuss the non-detection of the electronic C$_{60}$
transitions in the IC 418 optical spectrum and the possible connection between
fullerenes and diffuse bands, respectively. The conclusions of our work are
given in Section 7.


\section{Optical spectroscopy of PNe with fullerenes}

We acquired optical spectra of the fullerene PNe Tc 1 (B=11.1, E(B-V)=0.23;
Williams et al. 2008), M 1-20 (B=13.7, E(B-V)=0.80; Wang \& Liu 2007), and IC 418
(B=9.8, E(B-V)=0.23; Pottasch et al. 2004). The detection of fullerene-like
features in the IC 418 {\it Spitzer} spectrum has recently been reported by Morisset
et al. (2012). Tc 1 displays a fullerene-dominated spectrum with no clear signs
of PAHs, while M 1-20 and IC 418 also show weak PAH-like features (see, e.g.,
Garc\'{\i}a-Hern\'andez et al. 2010; Meixner et al. 1996). All PNe in our
sample also show unidentified broad dust emission features centred at
$\sim$9$-$13 and 25$-$35 $\mu$m (see, e.g., Garc\'{\i}a-Hern\'andez et al.
2012a). The effective temperature of PN IC 418 (T$_{eff}$=36700 K) is very
similar to the one in Tc 1 (T$_{eff}$=34060 K), while M 1-20 (T$_{eff}$=45880 K)
is among the fullerene PNe with the hottest central stars (Otsuka et al. 2014).
Our sample PNe display round or else elliptical morphologies: round (Tc 1) and
elliptical (IC 418, M 1-20) (see Figure 1 in Otsuka et al. 2014).   

The observations of Tc 1 and M 1-20 were carried out at the ESO VLT (Paranal,
Chile) with UVES in service mode between May and September 2011 (see
Garc\'{\i}a-Hern\'andez \& D\'{\i}az-Luis 2013 for more observational details).
We used the 2.4" slit centred at the central stars of the two PNe (see
Figure 1 in Otsuka et al. 2014) and following the parallactic angle. This
configuration should give a resolving power of $\sim$15000 from $\sim$3300 to
9400 \AA. However, from the O$_2$ telluric lines at $\sim$6970 \AA, we measure a
much higher resolving power of about 37000. The signal-to-noise ratio (S/N) (in
the final combined spectrum) in Tc 1 is very high ($\sim$300 at 4000 \AA~and
$>$300 at longer wavelengths), which permitted us to search for the expected
electronic transitions of neutral C$_{60}$ and both strong and weak DIBs
(Garc\'{\i}a-Hern\'andez \& D\'{\i}az-Luis 2013). In M 1-20, however, the final
S/N ($\sim$20 at 4000 \AA~and $>$30 at wavelengths longer than 6000 \AA)
was not high enough to search the relatively broad (and weak) C$_{60}$ features
around 4000 \AA\ or the weakest DIBs. 

\begin{table*}
\caption{Observational parameters of fullerene PNe and their comparison stars.}
\centering
\small\begin{tabular}{lccccclccccc}
\hline\hline
Object  & l & b    & E(B-V) & V$_{r}$ & Ref\tablefootmark{a} & Comparison star  & l & b  & E(B-V) & V$_{r}$  & Ref\tablefootmark{a}  \\
\hline
Tc 1         &   345.2375 & $-$08.8350    & 0.23 & $-$94.0        & 1, 2, 7  & HR 6334      &    350.829 &  4.285        & 0.42   &   7.0   &  6, 9\\
M 1-20       &     6.187  &     8.362     & 0.80 &   75.0         & 3, 4, 7  & HR 6716      &    7.162   &   $-$0.034    & 0.22   & 4.2     &  6, 10\\
IC 418       &   215.212  & $-$24.284    & 0.23 &   62.0         & 4, 5, 8  & HR 1890      &  208.177   &   $-$18.957   & 0.08   &  29.5   &  6, 10 \\
\hline
HD 204827\tablefootmark{b}    &   99.167   &   5.552    & 1.06 &  20.0    &  6, 11   &    &   & &   &   &     \\
\hline\hline
\end{tabular}
\tablefoot{
\\
\tablefoottext{a}{References. (1) Williams et al. (2008); (2) Frew et al. (2013); (3) Wang \& Liu (2007); (4) McNabb et al. (2013); (5) Pottasch et al. (2004); (6) Wegner
(2003); (7) Beaulieu et al. (1999); (8) Wilson (1953); (9) Kharchenko et al. (2007); (10) Pourbaix et al. (2004); (11) Petrie \& Pearce (1961).}
\tablefoottext{b}{The Hobbs et al. (2008) reddened star used as a reference for DIBs (see Section 3).}
}
\end{table*}

The optical spectroscopic observations of IC 418 (the brightest fullerene PNe in
our sample) were carried out at the Nordic Optical Telescope (NOT; Roque de los
Muchachos, La Palma) in March 2013 (under service time) with the FIES
spectrograph. The optical spectra were taken in the wavelength range
$\sim$3600-7200 \AA\ by using the FIES low-resolution mode (3630-7170 \AA;
orders 157-80) with the 2.5" fibre (centred at the IC 418 central star),
which translates into a resolving power of $\sim$25,000. Three exposures of 1200
s each were obtained in order to reach a S/N of $\sim$60 at 4000 \AA~(and in
excess of $\sim$150 at wavelengths longer than 5000 \AA) in the final combined
IC 418 spectrum.

As comparison stars, for Tc 1 and M 1-20 we selected the nearby B-type stars HR
6334 (B=5.1; E(B-V)=0.42; Wegner 2003) and HR 6716 (B=5.7; E(B-V)=0.22; Wegner
2003), respectively, while HR 1890 (B=6.4; E(B-V)=0.08; Wegner 2003) was selected
for IC 418. These comparison stars were observed on the same dates as the PNe
and with the same VLT/UVES and FIES set-ups. Two exposures of 300 s were enough
to obtain a final S/N in excess of $\sim$300 in the final combined spectra of
the comparison stars. The observed UVES and FIES spectra - processed with the
UVES data reduction pipeline (Ballester et al. 2000) and with the FIES reduction
software (FIEStool\footnote{See http://www.not.iac.es/instruments/fies/fiestool/FIEStool-manual-1.0.pdf}),
respectively - were corrected for heliocentric motion and combined, and the
stellar continuum for the three PNe was fitted by using standard astronomical
tasks in IRAF\footnote{Image Reduction and Analysis Facility (IRAF) software is
distributed by the National Optical Astronomy Observatories, which is operated
by the Association of Universities for Research in Astronomy, Inc., under
cooperative agreement with the National Science Foundation.}. Table 1 lists some
observational parameters, such as galactic coordinates, colour excess, and radial
velocity for the three fullerene PNe in our sample and their corresponding
comparison stars. 


\section{DIBs towards PNe}

We followed the list of DIBs measured in the high-S/N HD 204287 spectrum
(Hobbs et al. 2008) to search for them in the VLT/UVES spectra of Tc 1 and M 1-20, as
well as in the NOT/FIES spectrum of IC 418.

Our list of DIBs in Tc 1, M 1-20, and IC 418 are displayed in Tables A.1,
A.2, and A.3, respectively, where we give the measured central
wavelength ($\lambda$$_{c}$), the full width at half maximum (FWHM) as defined
in Hobbs et al. (2008), the equivalent width (EQW\footnote{One-sigma detection
limits for the EQWs in our spectra scale as $\sim$1.064  x FWHM / (S/N) (see,
e.g., Hobbs et al. 2008).}), the S/N in the neighbouring continuum, and the
normalized equivalent widths (EQW/E(B-V)). For comparison, we also list in
Table A.3 the EQW/E(B-V) values measured in HD 204827
and field-reddened stars by Hobbs et al. (2008) and Luna et al. (2008),
respectively. The DIB parameters were measured using standard tasks in IRAF
with no assumption on the DIB profiles; the only exception was the 4428 \AA\
DIB for which we assumed a Lorentzian profile (see, e.g., Snow et al. 2002). We
also list these parameters for the various interstellar components in those
DIBs that are clearly resolved (e.g., the 6196 and 6379 \AA~DIBs; Table A.1). 

\begin{figure*}
   \centering
   \includegraphics[angle=0,scale=.45]{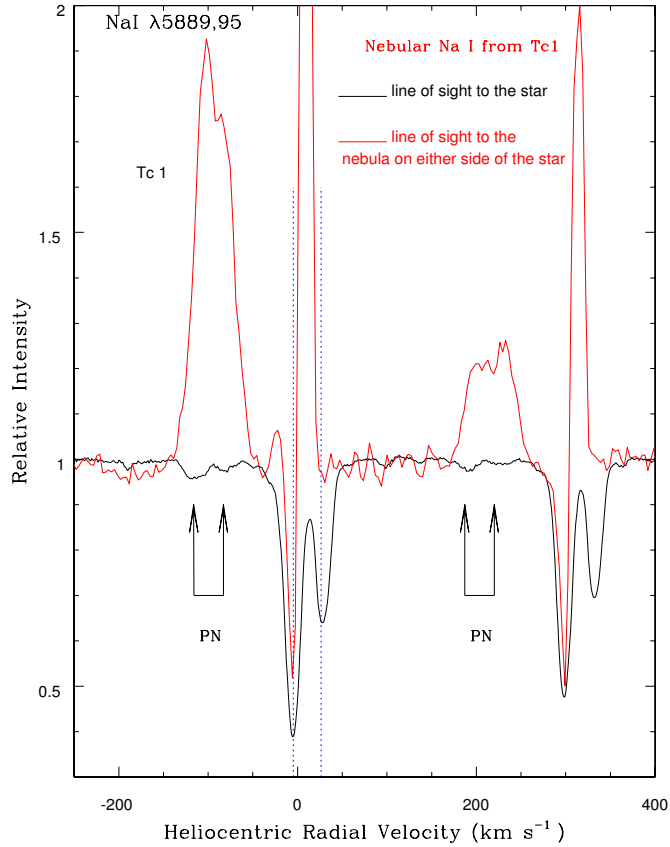}
   \caption{Na I D lines observed towards PN Tc 1 central star (black) and
average of two positions in the nebula 2.7 arcsec away from the central star on
either side (red) (from Williams et al. 2008). The dashed blue lines indicate
the interstellar Na I D components at $-$6.8 and $+$25 kms$^{-1}$. The sharp
emission close to 0 kms$^{-1}$ radial velocity is earth's airglow. The
absorption spectrum is in the stellar continuum units, while the emission
spectrum is in the nebular continuum units.
   \label{Fig1}}
   \end{figure*}

The Tc 1 optical spectrum displays two interstellar clouds at $-$6.80 kms$^{-1}$
and $+$25.00 kms$^{-1}$, as indicated by the two absorption components for the
atomic (Na I, Ca I) and molecular (CH$^{+}$) lines (see Figures 1 and 2, Table
A.4). Even the narrower DIBs towards Tc 1 (as 6379 \AA) show this behaviour (see
Figure 2 and Table A.4). The same behaviour is shown by the comparison star HR
6334, which also shows the two interstellar components at the same radial
velocities as Tc 1, confirming that both stars map similar ISM conditions. 

The M 1-20 optical spectrum, however, displays one Na I interstellar
component (i.e., at $-$6.54 kms$^{-1}$) with a blue assymetry that corresponds
to a weaker non-resolved Na I interstellar component. Both components for the
atomic Na I (e.g., at $-$6.12, and $-$26.43 kms$^{-1}$) interstellar lines are
clearly resolved in the comparison star HR 6716 (see Figure 3 and Table A.5),
with the peculariaty that the Na I lines are broader in M 1-20. Only the
strongest Na I interstellar component seems to be resolved in the DIBs (see
Figure 3). The presence of two clear (and narrower) interstellar components in
the comparison star suggests that both stars could map slightly different ISM
conditions. For M 1-20 and its comparison star, we list the DIB parameters for
the entire interstellar absorption (with no assumption on the DIB profile; Table
A.2).

The PN IC 418 seems to show also a main Na I interstellar component at a radial
velocity of $+$22.22 kms$^{-1}$ and a much weaker, not completely resolved
component at $\sim$$+$5 kms$^{-1}$ (see Figure 4 and Table A.6). Thus, the DIB
parameters for IC 418 are representative of the most prominent interstellar
component (Table A.3). In the comparison star HR 1890, two interstellar
components for the atomic Na I (e.g., at $+$5.42 and $+$23.75 kms$^{-1}$) are
clearly resolved, suggesting that both stars could map slightly different ISM
conditions. Anyway, we list the DIB parameters for the entire interstellar
absorption in both stars. We note that HR 1890 displays a very low reddening of
E(B-V)=0.08, and only the strongest DIBs are clearly detected. 

We identified 20, 12, and 11 DIBs in Tc 1, M 1-20, and IC 418,
respectively. All of these absorption bands are known DIBs, as previously
reported by Hobbs et al. (2008). It should be noted here that we could not
estimate the total absorption of the well studied 6993 and 7223 \AA\ DIBs in our
three PNe because of the strong meddling from the telluric lines. We also note
that our radial velocity analysis in Tc 1 shows that the $\sim$6309 and 6525
\AA~absorption features reported by Garc\'{\i}a-Hern\'andez \& D\'{\i}az-Luis
(2013) should be identified as stellar He II absorption lines\footnote{These are
the $\sim$6310 and $\sim$6527 \AA~stellar He II absorption lines, which are
blue-shifted by $\sim$90 kms$^{-1}$ (the central's star velocity) in Tc 1.}.
Their relative strengths (and widths) are consistent with the series of He II
features that are clearly seen in our Tc 1 optical spectrum (e.g., at 6171, 6234, 6406, 6683, 6891, 7178, 8237, and 9345 \AA). 


\subsection{Normal DIBs}

We concentrate here on those DIBs that seem to be normal for their reddening,
the so-called ``normal'' DIBs. In Tc 1, the PN with the highest quality spectrum
and a proper comparison star, their strengths are consistent with the E(B-V)
value, and both PN and its comparison star display similar normalized equivalent
widths (see Table A.1 and Figure 2). There are fifteen normal DIBs in Tc 1: six
of them are among the strongest DIBs most commonly found in the ISM (5797, 5850,
6196, 6270, 6379, and 6614 \AA)\footnote{The parameters for the 5797 and 5850
\AA~DIBs are more uncertain in Tc 1 because of their low intrinsic intensity and
some contamination by nearby spectral features. For example, there is a strong
stellar absorption line of C IV at 5799.84 \AA~and a nebular emission feature in
the proximity of the 5850 \AA~DIB.}, while the other nine DIBs (5776, 6250,
6376, 6597, 6661, 6792, 7828, 7833, and 8038 \AA) are  weaker interstellar
features already reported by Hobbs et al. (2008). The situation is less clear
for the fullerene PNe M 1-20 and IC 418. This is because the comparison stars
for both PNe seem to map slightly different ISM conditions (see above). The
comparison star of IC 418 also displays a very low reddening that prevents
detection of a significant number of DIBs. Despite this, the classification of
the latter DIBs (if detected in our spectra) as ``normal'' DIBs holds here for M
1-20 and IC 418 (see below).

The strengths of the sextet of common DIBs (5797, 5850, 6196, 6270, 6379, and
6614 \AA) are roughly consistent with the interstellar reddening in our three
PNe. The EQW/E(B-V) ratio of these DIBs for the three PNe agree reasonably well
with the values measured in their corresponding comparison stars (Tables A.1-3)
or with the EQW/E(B-V) values observed in HD 204827 (Hobbs et al. 2008) and/or
field-reddened stars (Luna et al. 2008). The 6196, 6376, 6379, and 6661 \AA~DIBs
towards Tc 1 (and its comparison star HR 6334) display the two interstellar
components at the same radial velocities in both objects (Table A.1 and Figure
2). In M 1-20 and its comparison star HR 6716, we could measure only one
main interstellar component for all DIBs (Table A.2 and Figure 3). As
in the case of M 1-20, all common DIBs towards IC 418 (and its comparison star
HR 1890) only display one main interstellar component (Table A.3 and Figure 4).
We note, however, that only a few DIBs (five) are detected towards HR 1890
(Table A.3), and all DIBs are intrinsically very weak due to the very low reddening
(E(B-V)=0.08) in the HR 1890 line of sight. Thus, Table A.3 also lists the
EQW/E(B-V) values of HD 204827 (Hobbs et al. 2008) and field-reddened stars
(Luna et al. 2008) for comparison with IC 418. The EQW/E(B-V) values of the
sextet of common DIBs in IC 418 are similar to those in HD 204827 and field-reddened stars. 

\begin{figure*}
   \centering
   \includegraphics[angle=0,scale=.45]{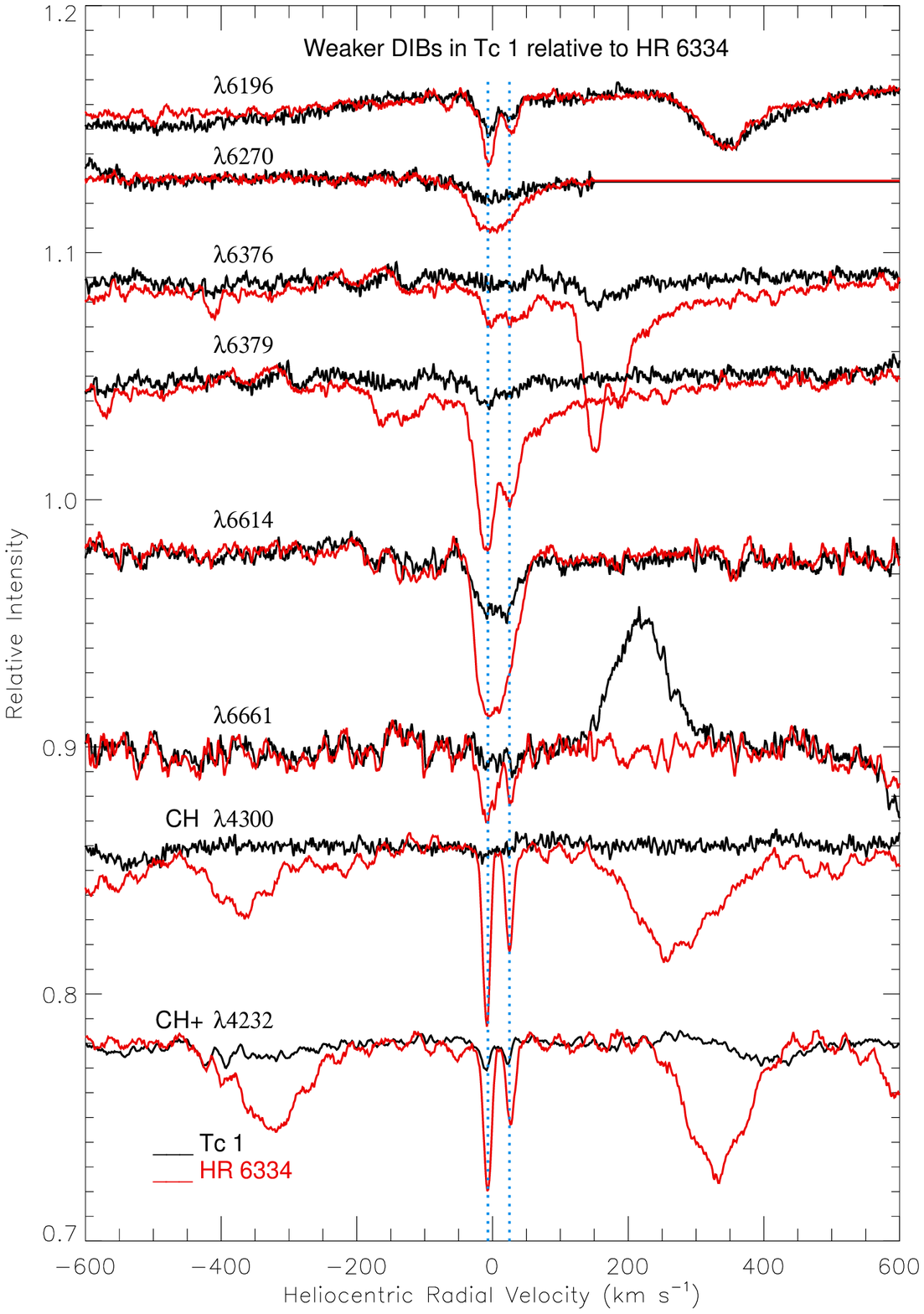}%
   \includegraphics[angle=0,scale=.45]{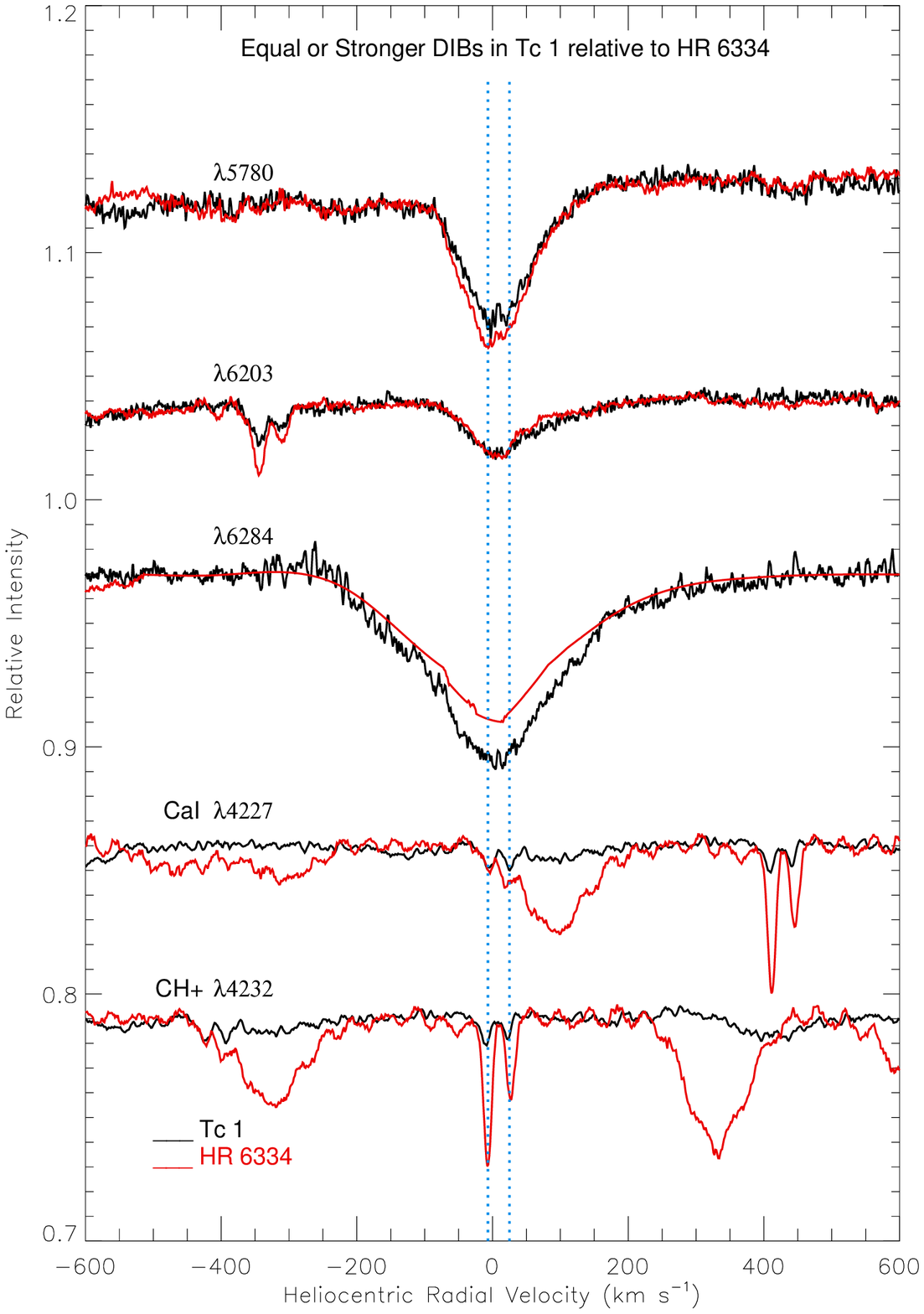}
    \caption{Profiles of a selection of DIBs with various strengths and widths are
displayed with respect to the heliocentric radial velocity. Weaker (or nearly
equal) and stronger DIBs in Tc 1 (in black) relative to HR 6334 (in red) are
displayed in the left and right panels, respectively. The profiles are shifted
vertically for clarity. The CH, CH$^{+}$, and Ca I profiles at the bottom show
the two interstellar components at $-$6.8 and $+$25 kms$^{-1}$ (marked with blue
vertical dotted lines) seen in both sources. The sharper DIBs in both
Tc 1 and HR 6334 also show these two components. The DIBs displayed in the left
panel are consistent with the lower reddening of E(BV)= 0.23 for Tc 1 as
compared to E(B-V)=0.42 for HR 6334. The right panel, however, shows that the
carrier(s) of these DIBs are enhanced (for the given E(B-V)) in the sight line
to Tc 1. \label{Fig2}}
    \end{figure*}

\begin{figure*}
   \centering
   \includegraphics[angle=0,scale=.45]{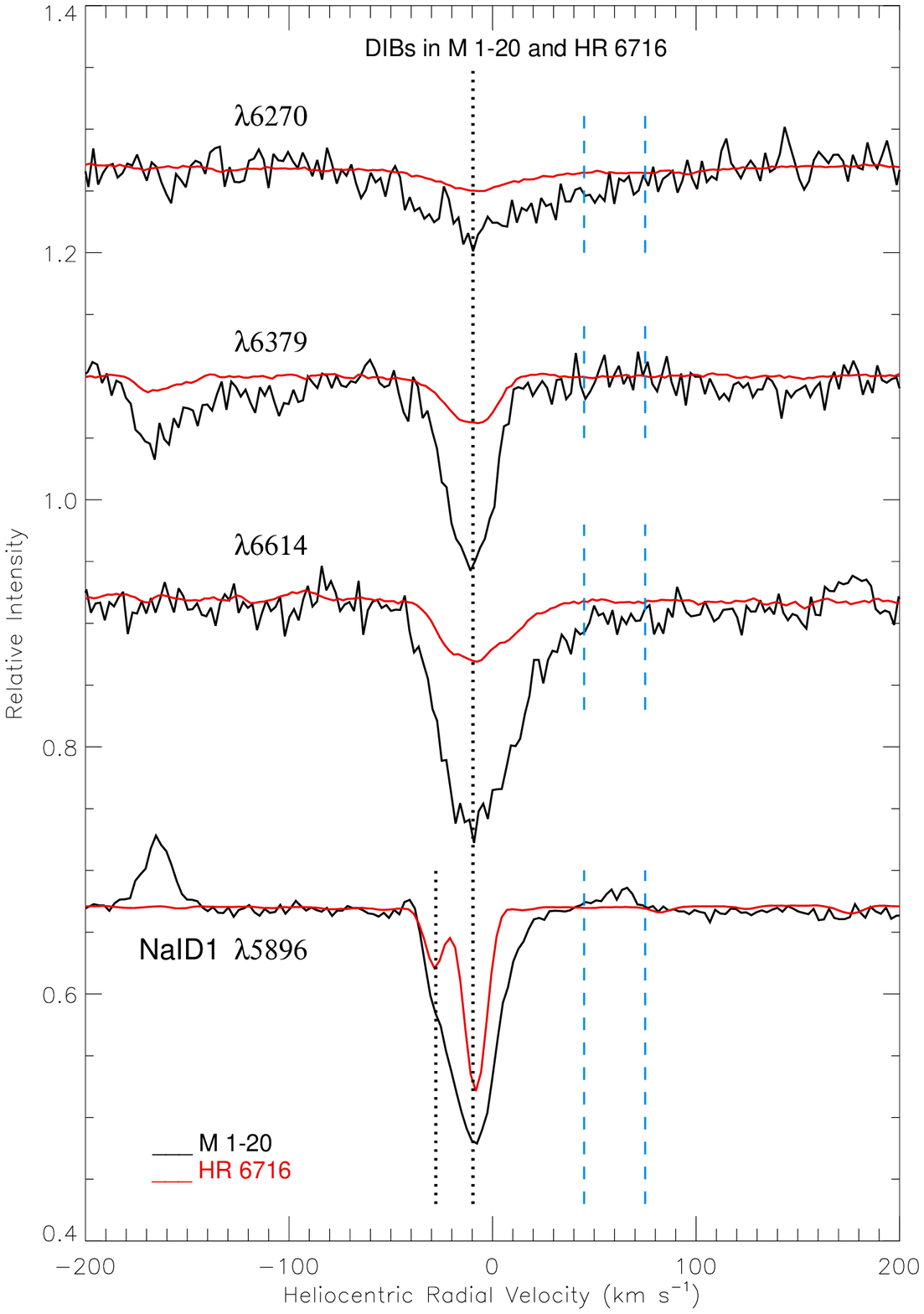}%
   \includegraphics[angle=0,scale=.45]{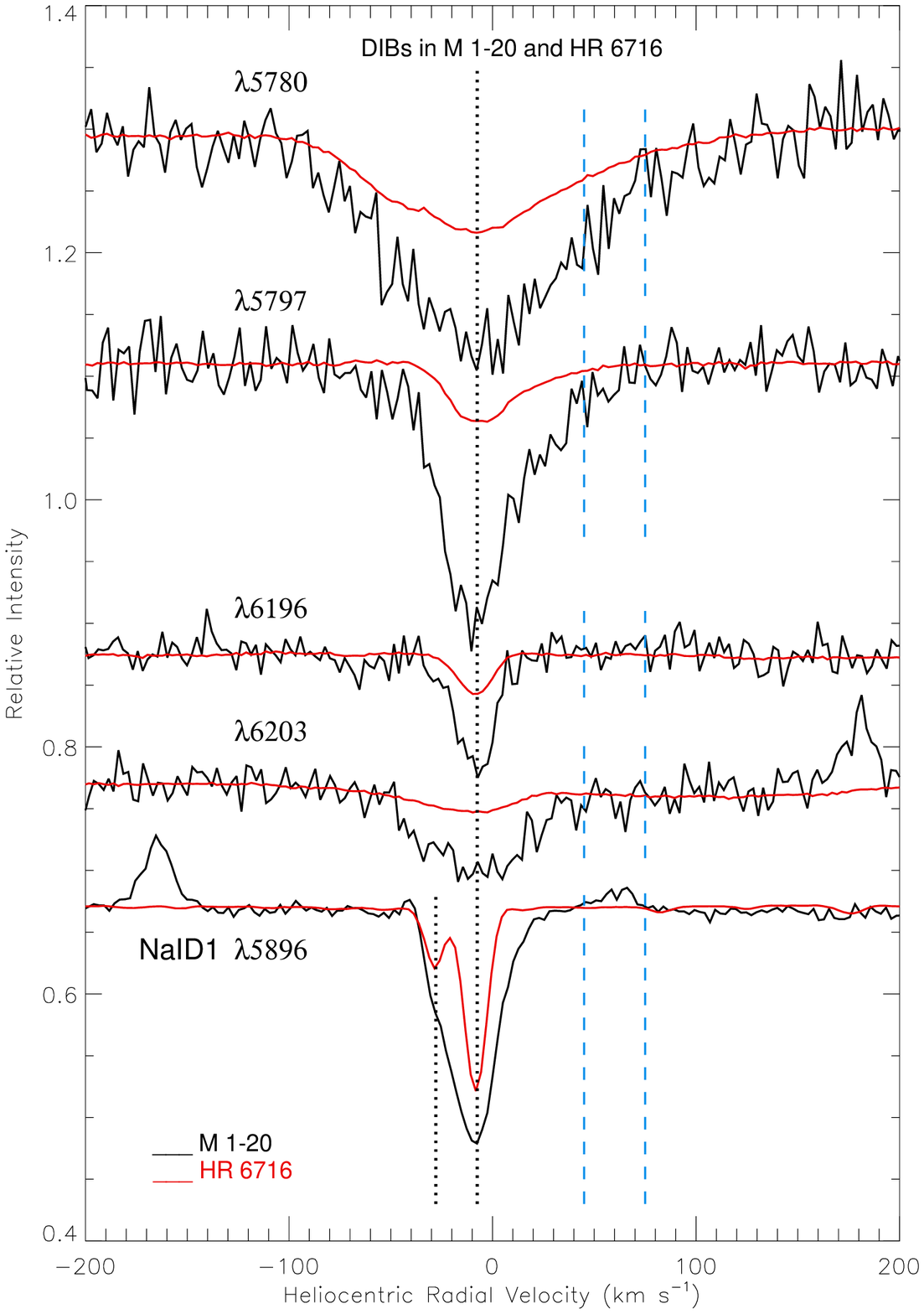}
    \caption{Profiles of a selection of DIBs with various strengths and widths with respect to the heliocentric radial velocity. DIBs with similar
and different profiles in M 1-20 (in black) relative to HR 6716 (in red) are
displayed in the left and right panels, respectively. The profiles are shifted
vertically for clarity. The Na I profiles at the bottom show the
interstellar components (marked with black vertical lines) seen in M 1-20 and HR
6716. The DIBs displayed in the two panels are consistent with the higher
reddening of E(BV)= 0.80 for M 1-20 as compared to E(B-V)=0.22 for HR 6716. The
stellar (and nebular) radial velocity range for M 1-20 is denoted by blue vertical
lines. \label{Fig3}}
    \end{figure*}

The measured intensities of the nine other ``normal'' DIBs (at $\sim$5776, 6250,
6376, 6597, 6661, 6792, 7828, 7833, and 8038 \AA)\footnote{The parameters for
the 5776 \AA~DIB are uncertain in Tc 1 because of its low intrinsic intensity
and contamination by a nearby nebular emission feature.} are also roughly
consistent with the E(B-V) values in our three fullerene PNe. All these DIBs are
detected in Tc 1. For the narrower 6196, 6376, 6379, and 6661 \AA~DIBs in Tc 1
and its comparison star, we also give the parameters for each one of the
interstellar components mentioned above (Table A.1). In M 1-20, however, only
two (6376 and 6661 \AA) of these weak DIBs are detected in our spectrum
(Table A.2). In IC 418, we only detect the weak DIB at 6376 \AA\ because the
other weak DIBs are below our 1-$\sigma$ detection limits.

In short, the carriers of the so-called ``normal'' DIBs do not seem to be
particularly over-abundant towards fullerene PNe, since they are consistent with those
expected for the general diffuse ISM.

          
\subsection{Unusually strong DIBs}

Interestingly, some DIBs are found to be unusually strong in fullerene PNe. The 
five DIBs at $\sim$4428, 5780, 6203, 6284, and 8621 \AA\ are unusually strong
towards Tc 1. Their strengths in Tc 1 are higher than expected for the E(B-V) of
0.23; Tc 1 displays EQW/E(B-V) values higher than those in the comparison star
HR 6334 (Table A.1 and Figure 2). This is clearly shown in Figure 5, where we
plot the EQW/E(B-V) values of DIBs in Tc 1 versus HR 6334 (left panel) and those
in the reference star HD 204827 versus HR 6334 (right panel). The EQW/E(B-V)
values of DIBs in HR 6334 scale nicely with those in HD 204827 (the only
exception is the 6284 \AA~DIB), suggesting similar ISM properties towards both
stars. In Tc 1, however, the five unusually strong DIBs mentioned above clearly
deviate from the linear relation followed by most of the DIBs that also scale
well with those in HR 6334 (and HD 204827). The central radial velocity of these
unusually strong interstellar features is the same in both Tc 1 and the comparison
star (with the apparent exception of the 4428 \AA\ feature; see Section 4),
confirming their interstellar origin. In addition, the neutral molecular lines
(CH, CN) are much weaker or absent towards Tc 1 (Figure 2), indicating a higher
degree of ionization. This may indicate that carriers of these DIBs (enhanced
towards Tc 1) may be ionized species.

The situation is again less clear for the fullerene PNe M 1-20 and IC 418. It
seems clear, however, that at least the $\sim$4428 \AA\ DIB is unusually strong
in both PNe (see below), showing EQWs that are much higher than expected for their E(B-V)
values. 

For the well-studied 4428 \AA~DB we adopted a Lorentzian profile (Snow et al.
2002), obtaining EQWs of $\sim$860, 2579, and 1001 m\AA~for Tc 1, M 1-20, and IC
418, respectively. The 4428 \AA~DB in Tc 1 and IC 418 is at least a factor of
two greater than expected for their low reddening of E(B-V)=0.23 (see e.g., Fig.
6 and 15 in Snow et al. 2002 and van Loon et al. 2013, respectively), while this
DB is $\sim$1.5 times more intense than expected in M 1-20.  

The DIB at 5780 \AA\ is another intereresting feature that also could be an
unusually strong DIB towards IC 418 (Table A.3). In IC 418, it is stronger
(EQW/E(B-V)=0.43) than in the comparison star HR 1890 (with EQW/E(B-V)=0.32) and
in the Hobbs et al. (2008) reference star HD 294827. However, it has a similar strength
to field-reddened stars (Luna et al. 2008). Unfortunately, the DIBs at
$\sim$6203 and 6284 \AA\ are not present in the low-reddening star HR 1890.
Similar to the 5780 \AA\ DIB, these DIBs in IC 418 are stronger than in the
star HD 294827 but of similar strenght to those seen in the sample of
field-reddened stars by Luna et al. (2008). Finally, our IC 418 optical spectra
do not cover the spectral region around the 8621 \AA\ feature.

As mentioned above, the situation is less clear also for M 1-20. Indeed,
the 5780 and 6284 \AA~DIBs in M 1-20 are weaker than in HR 6716, while the 6203
\AA~DIB is of equal strength in both sources. Apparently, the bands at 5780,
6203, and 6284 \AA~seem to be unusually strong towards both M 1-20 and HR 6716
(see Table A2). As in the case of IC 418, these DIBs are stronger than in the
reference star HD 294827 but similar to field-reddened stars. The most
remarkable outcome is the complete lack of the 8621 \AA\ DIB towards M 1-20, which
otherwise is very intense in the Tc 1 line of sight.

\begin{figure*}
   \centering
   \includegraphics[angle=0,scale=.45]{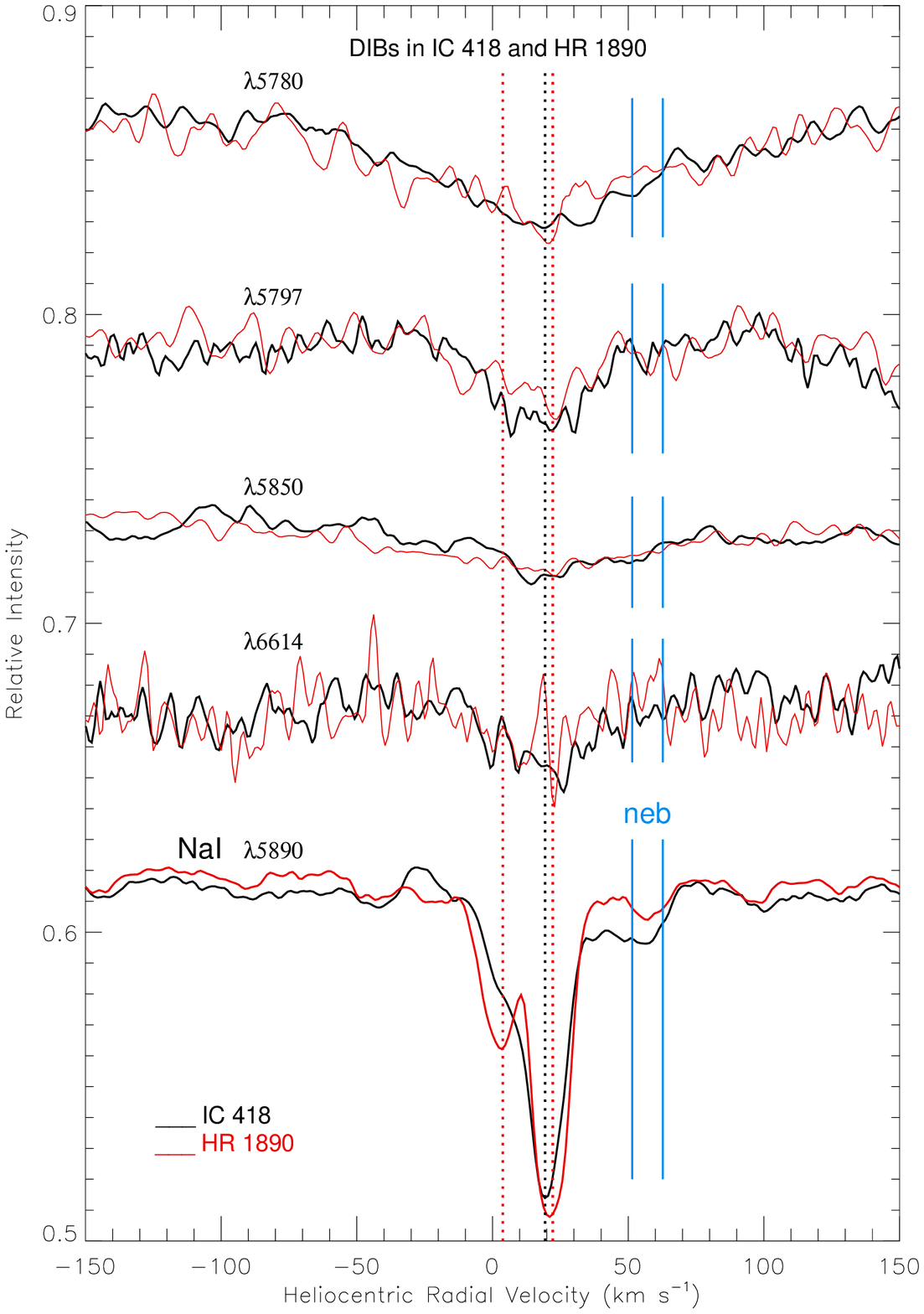}%
   \includegraphics[angle=0,scale=.45]{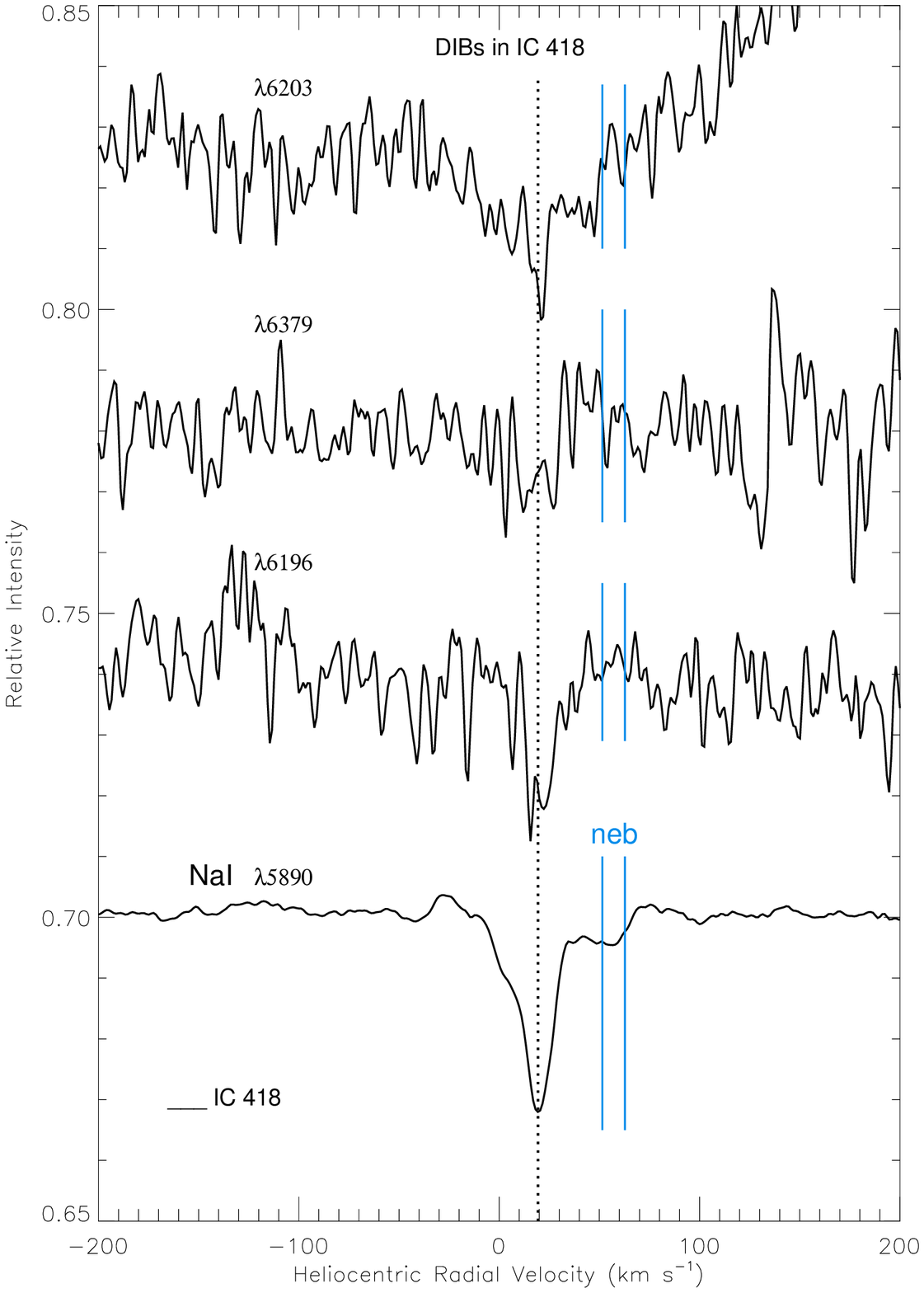} 
    \caption{Profiles of a selection of DIBs with various strengths and widths are
displayed with respect to the heliocentric radial velocity. Some normal and
unusually strong DIBs in IC 418 are displayed (not in HR 1890 due to the low
reddening of 0.08). The Na I profiles show the prominent component at 22.22
kms$^{-1}$ (marked with a black vertical dotted line) in IC 418.
\label{Fig4}}
    \end{figure*}
    
\begin{figure*}
   \centering
   \includegraphics[angle=0,scale=.45]{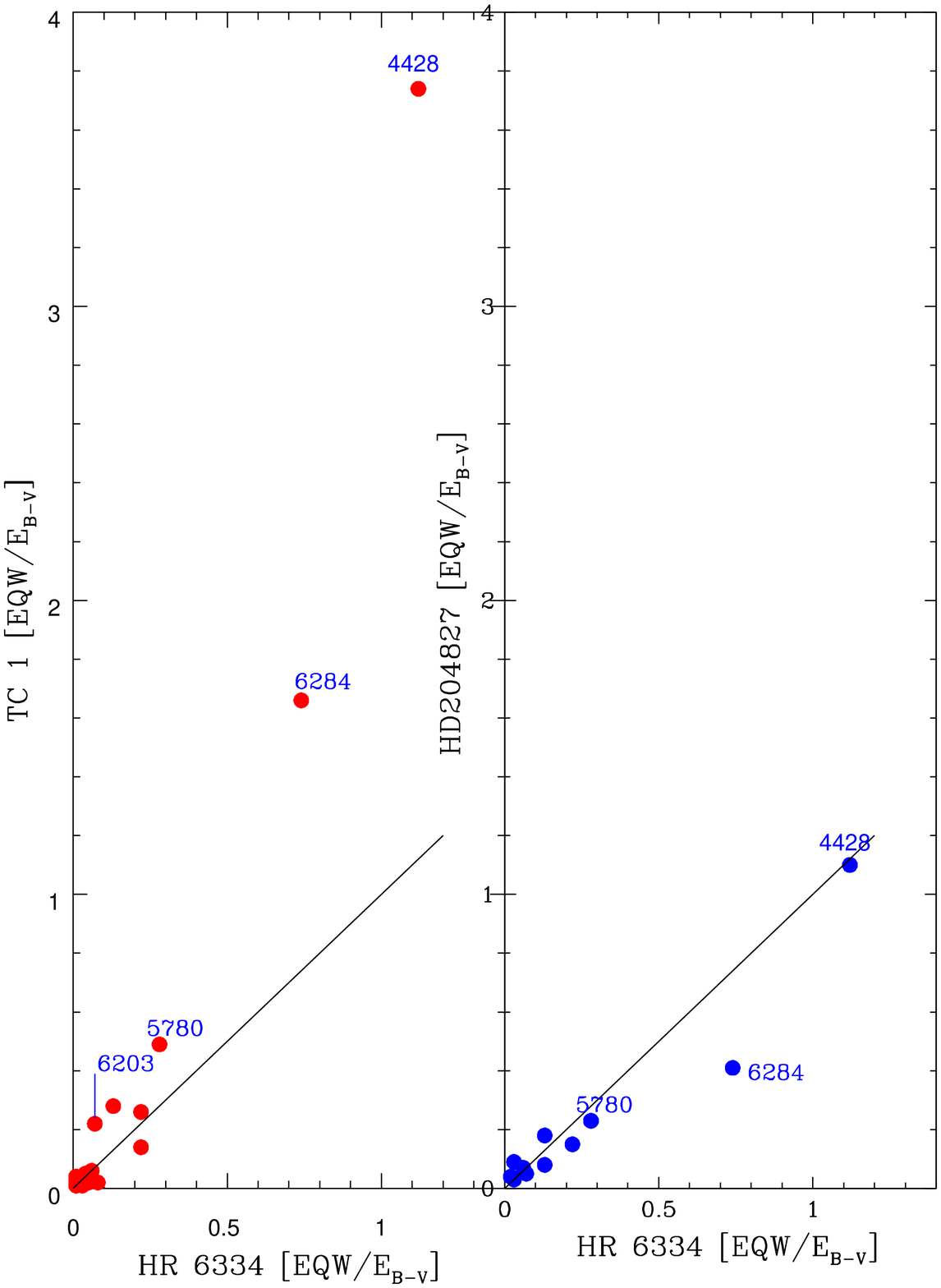}
   \caption{Plots of EQW/E(B-V) of Tc 1 with respect to (w.r.t.) HR 6334 (left
panel) and HD 204827 w.r.t. HR 6334 (right panel). The EQW/E(B-V)
values of HR 6334 scale nicely w.r.t. HD 204827, which suggests that the
properties of ISM are similar for DIBs towards the two stars. For Tc
1, most of the DIBs also scale well w.r.t. HR 6334 (and HD 204827) with the
exception of the five unusually strong DIBs (those at $\sim$4428, 5780, 6203,
6284, and 8621 \AA; see text), which deviate from the linear relation.
\label{Fig5}}
    \end{figure*}
    
Finally, it is worth mentioning here that the DIB at $\sim$6203 \AA\
varies among the fullerene PNe in our sample. This DIB and the other band
at 6205 \AA~are usually measured as two distinct interstellar absorptions
(e.g., corresponding to different carriers; see, e.g., Porceddu et al. 1991). The
complex band at $\sim$6203 \AA~is especially noteworthy, in which the EQW/E(B-V)
ratio is two to three times higher for Tc 1 than for the comparison star HR 6334 and the
reddened star HD 204827 (Hobbs et al. 2008). Towards IC 418, the 6203 \AA~DIB is
not very strong, although is not detected in the line of sight of the low-reddening comparison star HR 1890. The 6203 \AA~DIB towards M 1-20 and its
comparison star HR 6716 displays a similar EQW/E(B-V) ratio. Curiously, the
secondary DIB at $\sim$6205 \AA~is not clearly detected towards any of our
sample PNe. This DIB is not easily recognized (resolved) from the dominant 6203
\AA~DIB. Furthermore, we can find no evidence for the presence of the 6205
\AA~interstellar absorption in IC 418 because it coincides with a strong nebular
emission line.   


\section{A search for diffuse circumstellar bands}

The detection of diffuse circumstellar bands (DCBs) is very difficult because
high S/N spectra are mandatory for detecting the presumably much weaker
circumstellar features. Also, the DCBs have to be distinguished from the DIBs,
and this distinction can only be made by measuring the radial velocities of
the circumstellar and interstellar components in the line of sight to our PNe.
The S/N is too low in PN M 1-20, preventing any search for DCBs towards this
object. A higher quality (S/N$\sim$100-200) optical spectrum was obtained for
IC 418, but its relatively low radial velocity ($\sim$58 kms$^{-1}$; Dinerstein
et al. 1995) makes it difficult to distinguish possible DCBs from the DIBs.
However, our high-quality (S/N $>$ 300) spectra for PN Tc 1, together with its
higher radial velocity (in the range from -83 to -130 kms$^{-1}$; Williams et
al. 2008), may permit us to search for the possible presence of DCBs. 

\begin{figure*}
   \centering
\includegraphics[angle=0,scale=.45]{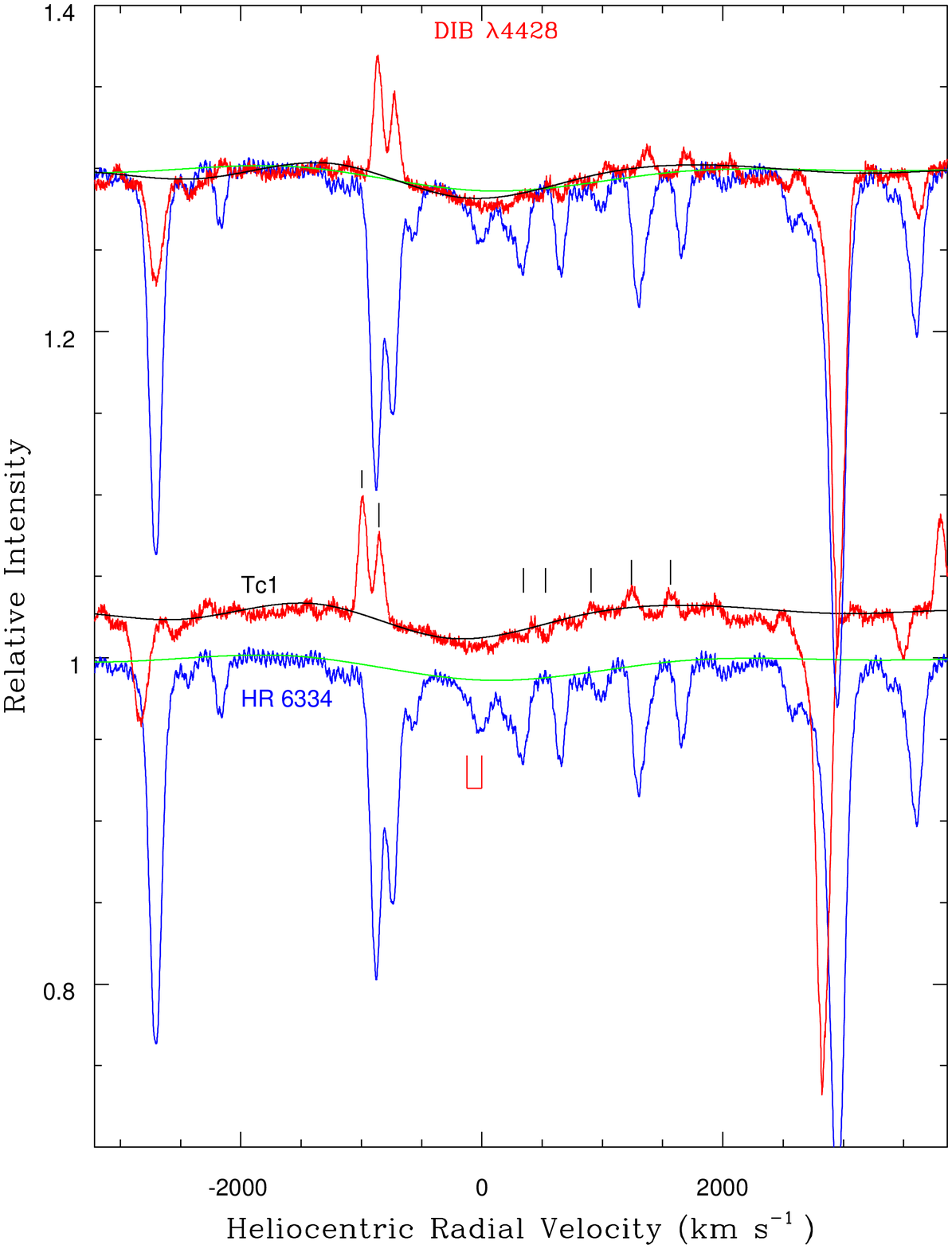}
\caption{Profiles of the 4428 \AA~feature in Tc 1 (red) and in the comparison
star HR 6334 (blue). The black and green profiles of the 4428 \AA\ feature have
been constructed by avoiding the stellar emission and absorption lines (marked
by short black lines) and by fitting a high-degree polynomial function to the
clearer regions. The minimum of the 4428 \AA\ feature in HR 6334 occurs around
0.0 kms$^{-1}$, while it seems to be blue-shifted in Tc 1. At the top, the
profiles of Tc 1 has been shifted by 126 kms$^{-1}$ redwards and superposed on
the HR 6334 profile to illustrate the apparent non-coincidence of the minima of
the profiles in both stars. \label{fig6}}
\end{figure*}

\begin{figure*}
   \centering
   \includegraphics[angle=0,scale=.45]{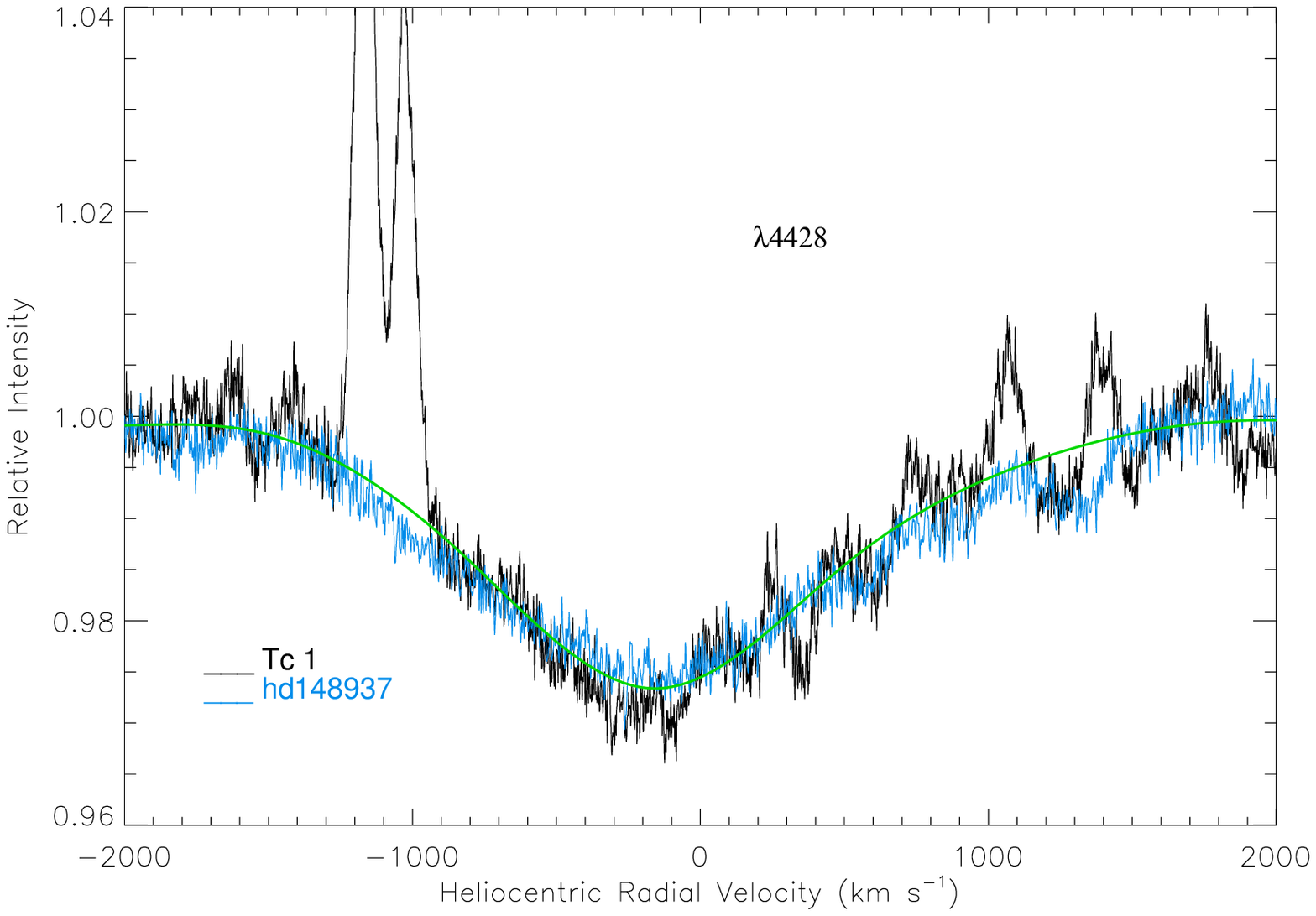}%
   \includegraphics[angle=0,scale=.45]{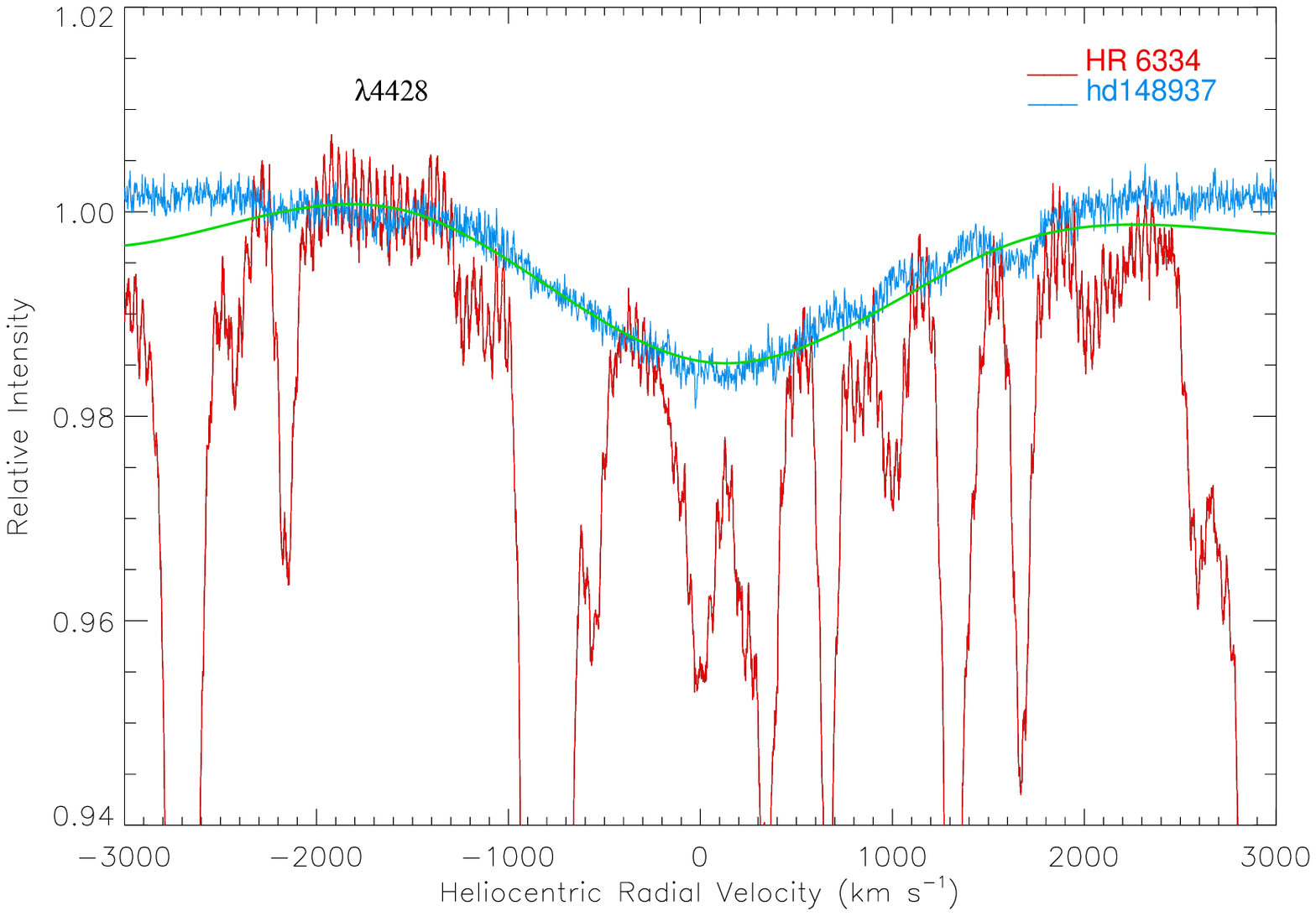} 
    \caption{Profiles of the 4428 \AA~feature in Tc 1 (left panel) and in
the comparison star HR 6334 (right panel) superposed on the profile of the O6.5
star HD 148937. The O6.5 star has been displaced (in velocity) in both
panels to make the 4428 \AA\ band coincide with the profile and the minima of
the 4428 \AA\ band towards Tc 1 and HR 6334. All spectra have
been normalized on the EQW of the 4428 \AA~band. The minimum of the feature in
HR 6334 occurs around 0.0 kms$^{-1}$, while it seems to be blue-shifted in Tc 1.
The green profiles of the 4428 \AA~feature in Tc 1 and HR 6334 are the
polynomial fits shown in Figure 6. The superposed spectrum of an O6.5 star
(in blue) support the non-coincidence of the profiles minima in Tc 1 and HR 6334.
\label{Fig7}}
    \end{figure*}

The nebular absorption lines in the line of sight to Tc 1 show heliocentric
radial velocities in the range of $-$83 to $-$130 kms$^{-1}$ (Williams et al.
2008), while the star's heliocentric radial velocity is measured as
$-$90$\pm$12 kms$^{-1}$. Any absorption/emission feature in the radial
velocity range from -80 to -130 kms$^{-1}$ is then expected to be related to
the Tc 1's expanding circumstellar (nebular) gas. Indeed, the Na I D lines
show two distinct absorption components at $-$83 and $-$116 kms$^{-1}$ apart
from the ISM components (see Figure 1 and Table A.4). The 5780
\AA~DIB towards Tc 1 matches the velocity of the ISM, as expected (see Figure
8). However, we find a (blue-shifted) very weak absorption feature (at the
$\sim$3-sigma level; EQW$\sim$6.8 m\AA\ and FWHM$\sim$0.99 \AA) at the Tc 1
nebular velocity (centred at $\sim$$-$125 kms$^{-1}$ and indiccated in Figure 8), which is not present in the comparison star HR 6334 (see
Figure 2). There is no known DIB at this wavelength, and it does not correspond to
any stellar line (too narrow) or telluric feature (not present in HR 6334). In
addition, the velocity separation is not in the right direction for another
ISM cloud. This circumstellar absorption feature is narrower than the
interstellar one, and the primary interstellar feature is broader owing to the
contribution of at least two interstellar clouds (see above). The
physical/chemical conditions in the Tc 1's circumstellar envelope are also expected
to be different from those in the ISM, and the widths may not be necesarily
the same. We note that there is a nebular emission counterpart (Figure 8) of the very weak DCB around 5780 \AA, and the
presence of this nebular emission at the radial velocity of Tc 1 furthermore
suggests its circumstellar origin (see below).

Curiously, the 4428 \AA\ feature in Tc 1 seems to be blue-shifted (by
$\sim$126 kms$^{-1}$) relative to the one in HR 6334 (see Figure 6). At the top
of Figure 6, the profiles of Tc 1 have been shifted by 126 kms$^{-1}$ redwards
and superposed on the HR 6334 profile to show the apparent difference in the
profiles minima in both stars. It also aligns the photospheric lines of both
stars, suggesting that the velocity of the 4428 \AA~absorption feature is
close to the Tc 1's nebular velocity. In Figure 7, the profiles (normalized on
the EQW of the 4428 \AA\ band) of Tc 1 (left panel) and HR 6334 (right panel)
have been superposed on the profile of an O6.5 star (HD 148937). This O6.5
star has been displaced (in velocity) in both panels to match the 4428 \AA\
profiles observed in both Tc 1 and HR 6334. The 4428 \AA~profiles both in Tc 1
and HR 6334 match the one in the O6.5 star (HD 148937) as well as our
polynomial fits (the green profiles in Figure 7)\footnote{Despite the
pollution by many lines in HR 6334, Figure 7 shows that the 4428 \AA~profile
in HR 6334 match our polynomial fit very well (the green profile in Figure 7
(right panel).} This indicates that the minima of the 4428 \AA~feature in Tc 1
and HR 6334 are well determined. The minimum of the 4428 \AA~feature in HR
6334 occurs around 0.0 kms$^{-1}$, while it is blue-shifted by $\sim$126
kms$^{-1}$ in Tc 1. The blue shift could either be a result of
vibrational-rotational structure of the carrier molecule or could be due to
radial motions of the carriers. The velocity shift is about the same amount as
the radial velocity of the circumstellar gas of Tc 1 (also the Na I D
absorption components). A nebular emission feature is present at the
wavelength corresponding to the blue shift of $\lambda$4428 absorption feature
(see below), which cannot be identified with any nebular line (see, e.g.,
Sharpee et al. 2003 for a complete compilation of nebular lines in PN IC 418
that displays an effective temperature that is almost identical to Tc 1). Thus, it
seems likely that the apparent blue shift of the 4428 \AA\ feature is real,
indicating a circumstellar (nebular) nature for the carrier(s). Also,
the ISM contribution to the 4428 \AA\ feature in Tc 1 is expected to be a minor
one; even smaller than the one in HR 6334.

\begin{figure*}
   \centering
   \includegraphics[angle=0,scale=.45]{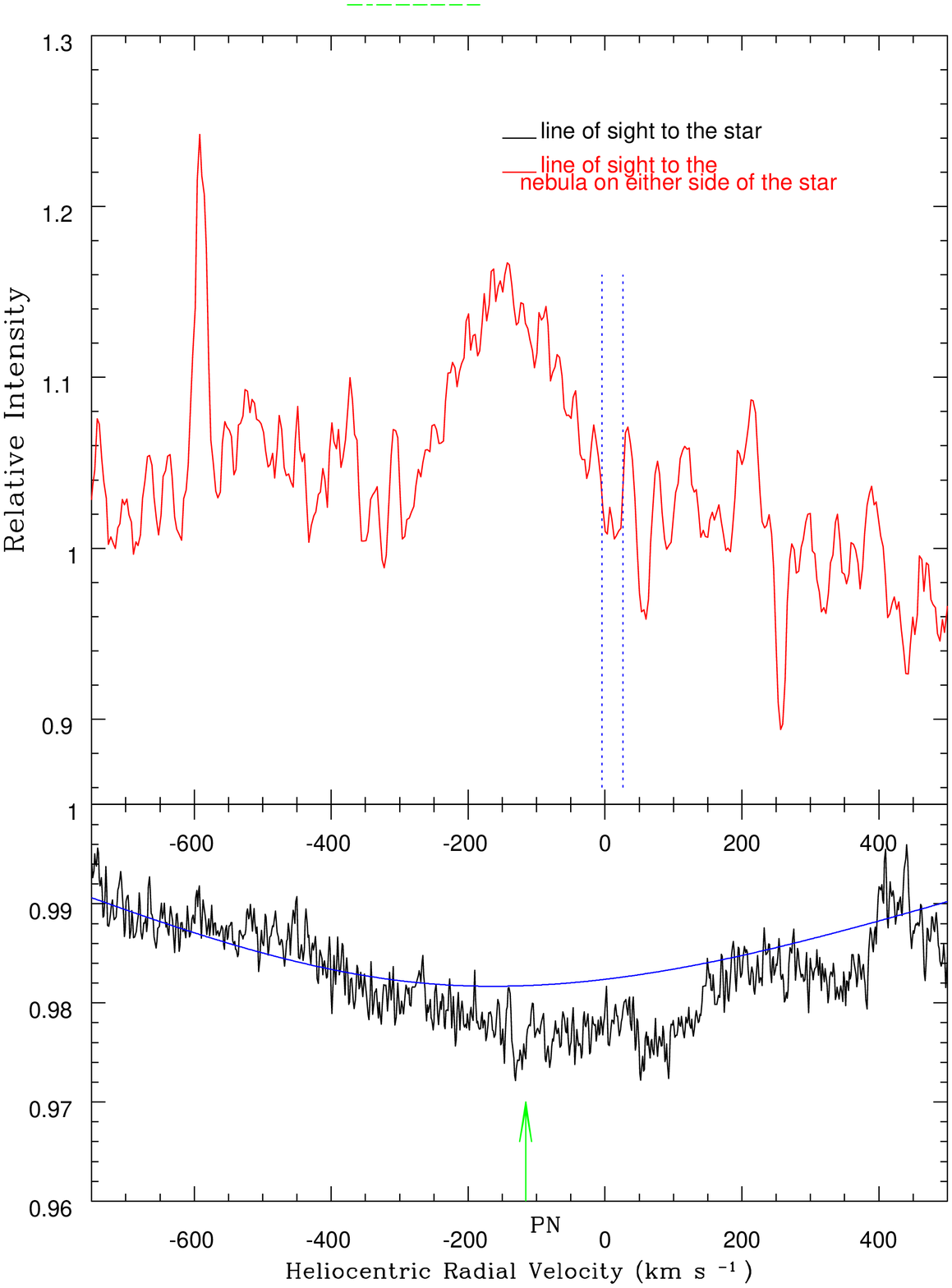}%
   \includegraphics[angle=0,scale=.45]{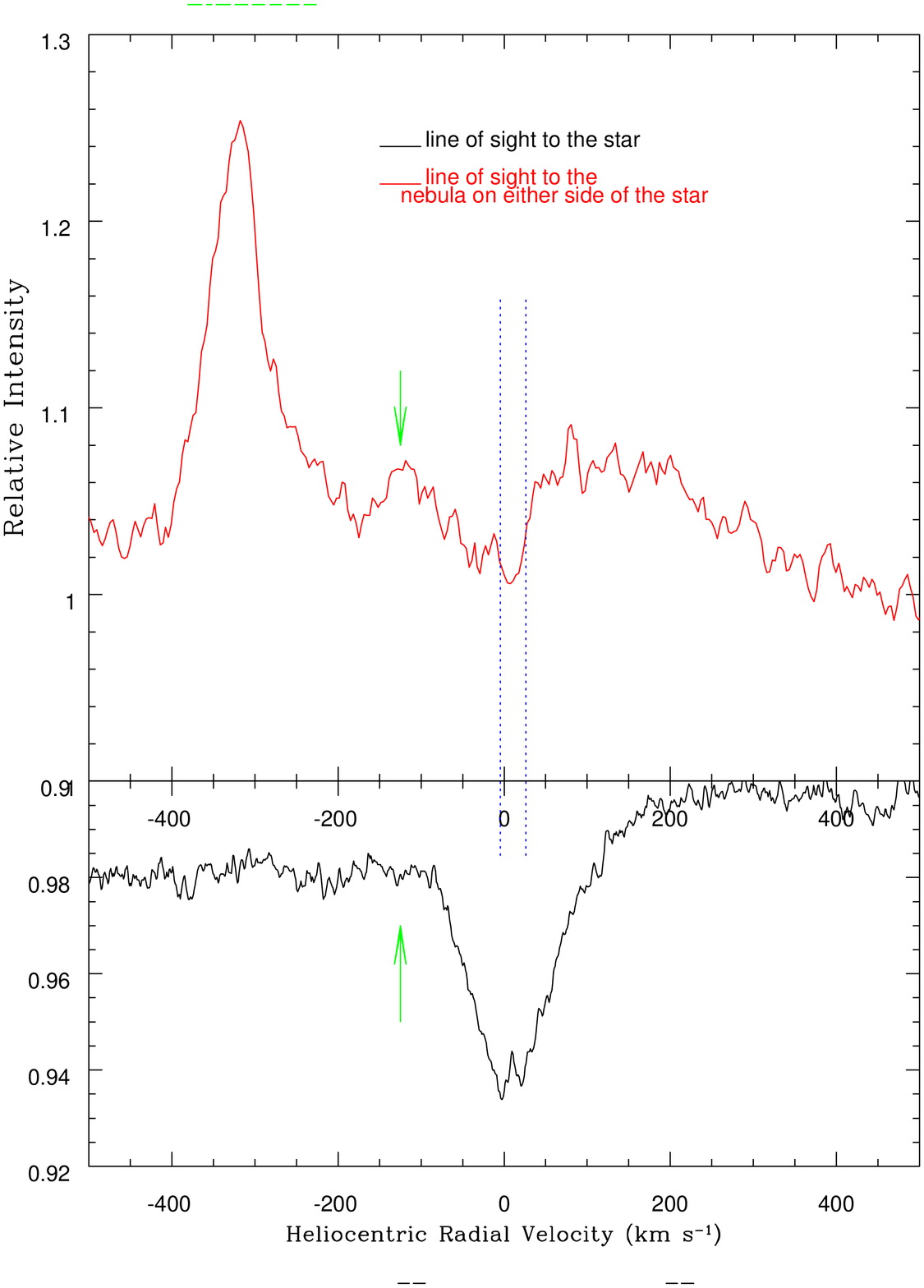} 
    \caption{Profiles of the broad 4428 \AA~band (left panel) and of the 5780
\AA~feature (right panel) towards Tc 1 central star (black) and average of two
sight lines to the nebular position on either side of the nebula (from
Williams et al. 2008). In both panels, the dashed blue lines mark the
interstellar components at $-$6.8 and $+$25 kms$^{-1}$. Note the coincidence
in velocity (marked by green arrows) of the profile centre of the broad 4428
and of the weak 5780 \AA~circumstellar absorptions and the corresponding
nebular emissions. The emission feature to the left of the 5780 \AA~nebular
emission is unidentified. \label{Fig8}}
    \end{figure*}

If the material in and around the nebula (plus central star) is giving rise to
the circumstellar absorption components in the sight line towards the star (as
seen in Na I D and some DIBs), then the same material is expected to be seen
in emission in the sight lines of the nebula away from the central star. We
have investigated the Tc 1 nebular spectra obtained by Williams et al. (2008)
and their spectroscopic observations in sight lines away from the central star
indeed seem to confirm this expectation. The average spectrum of the nebula at
two slit positions 2.7 arcsec away from the Tc 1 central star (Williams et al.
2008) - which samples the same {\it Spitzer} volume that revealed the
fullerenes in Tc 1 - shows the Na I D lines in emission (also the Ca II K
lines, but these are much weaker) at the radial velocity of the object (see
Figure 1 and Table A.1). The Na I D emission components have
slightly more positive radial velocity ($\sim$10 kms$^{-1}$ with respect to
the absorption lines), suggesting a possible expansion velocity of $-$10
kms$^{-1}$ for the Na I gas. Such a correspondence of emission feature occurs
with $\lambda$4428 absorption. Similar emission corresponding to $\lambda$5780
also seems to be present in the Tc 1 nebular spectrum. The presence of 4428
and 5780 \AA\ nebular emission (see Figure 8) at the radial velocity of Tc 1
furthermore suggests their circumstellar origin. The DCBs reported here are known
to be among the strongest DIBs. The strong 4428 \AA\ feature is known to
correlate well with other DIBs like 5780 (e.g., van Loon et al. 2013). The
5780 \AA\ DIB is also known to correlate with the broad 6284 \AA~feature in
the ISM (see, e.g., Friedman et al. 2011) but might differ in circumstellar
environments. Interestingly, the 6284 \AA\ DIB is very strong in Tc 1 and
might even be hiding a circumstellar absorption feature as well; in Figure 2,
there is evidence of some asymmetry in the 6284\AA~DIB profile at the Tc 1's
radial velocity range (from $-$83 to $-$130 kms$^{-1}$). 

The PN IC 418 also shows circumstellar (nebular) absorption components
(although much weaker than in Tc 1) in the Na I D lines (Dinerstein et al.
1995; see also Figure 4 and Table A.6). Indeed, the 5780
\AA~DIB towards IC 418 displays a tentative weak asymmetry (even weaker than
towards Tc 1) at the nebular velocity of $\sim$58 kms$^{-1}$. Unfortunately,
the 4428 \AA~feature is not seen in its (low reddening) comparison star HR
1890, and we could not properly check whether the profile minimum of this feature is
red-shifted to the observed nebular velocity of IC 418. The comparison of the
4428 \AA~profile in IC 418 with the one in the O6.5 star HD 148937 displayed
in Figure 9 tentatively suggests that this feature in IC 418 could be slightly
red-shifted with respect to the expected interstellar wavelength. However, the
4428 \AA~profile in IC 418 is not fully reproduced by the one in HD 148937, and
the central wavelength of this feature is uncertain (i.e., more uncertain than
in the case of Tc 1 above).

\begin{figure*}
   \centering
   \includegraphics[angle=0,scale=.45]{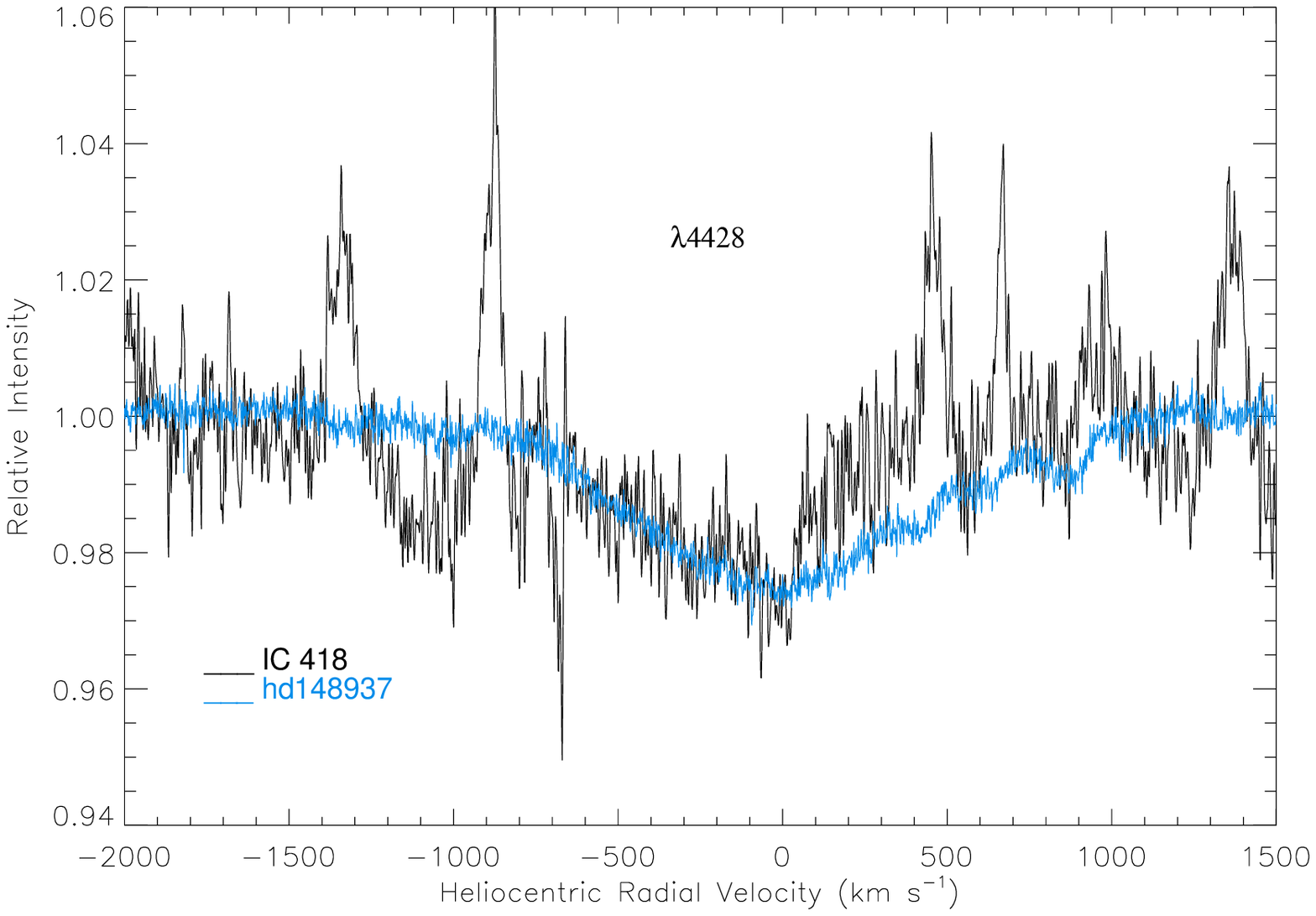}
    \caption{Profiles of the 4428 \AA~feature in IC 418 (in black) superposed on
the profile of the O6.5 star HD 148937 (in blue). The O6.5 star has been
displaced (in velocity) to try the 4428 \AA\ band coincide with the profile and
the minima of the 4428 \AA\ band towards IC 418. Note that both spectra have
been normalized on the EQW of the 4428 \AA~band. The 4428 \AA~profile in IC 418
is not fully reproduced by that in HD 148937 and its central wavelength is quite
uncertain (see text).\label{Fig9}}
    \end{figure*}

As mentioned above, the S/N in the M 1-20 spectrum is probably too low
for detecting DCBs in this object. The heliocentric radial velocity of
the nebula has been measured by us as $\sim$61 kms$^{-1}$ (from a few He I
nebular emission lines). This radial velocity is shown in Figure 3. In our low S/N M 1-20 spectrum, there is no
evidence of any DIB component close to this radial velocity.

Based on the EQW of the strong DIB at 5780 \AA\ towards Tc 1 (EQW=112.1 m\AA)
relative to the weak DCB around the same wavelength (EQW=6.8 m\AA), we can
estimate the S/N needed to detect this DCB in M 1-20 and IC 418. An
EQW(DIB)/EQW(DCB) ratio of 16.5 is obtained for Tc 1. Adopting this value for
the other PNe, one can estimate the expected EQW of the 5780 \AA~DCB; EQW(DCB)
values of 20.1, and 6.1 m\AA\ are obtained for M 1-20 and IC 418, respectively.
Then, when assuming the same FWHM (990 m\AA) as in Tc 1, the needed S/N for the DCB
feature to be detected at three sigma can be obtained (EQW$\sim$ 3  x FWHM / (S/N);
see Hobbs et al. 2008). We find that we would need S/N$\sim$143 and 491 for M
1-20 and IC 418, respectively. Thus, our non-detection of the DCBs in the latter
PNe is due to the lower S/N in our spectra for both objects (see Tables A.2 and
A.3).

In summary, the present data show that DCBs might not be uncommon in
fullerene-containing PNe and suggest the first detection of two DCBs at 4428
and 5780 \AA~in the fullerene-rich circumstellar environment around the PN Tc
1. However, we prefer to be cautious until these posible DCB detections are
confirmed in other PNe with fullerenes. The three fullerene-containing PNe in
our sample display very weak circumstellar absorptions of Na I, and the
intrinsic weakness of the DCBs (e.g., 5780 \AA) is very likely related with
the low column density of the gas (and dust) in their circumstellar envelopes.
The strength of the 5780 \AA~circumstellar absorption in fullerene PNe is
likely to be correlated with the circumstellar Na column density and the best
PNe to unambiguously confirm that our detection of DCBs are those showing a strong
Na I circumstellar absorption that is well separated (i.e., at a very different radial
velocity) from the Na I interstellar components.


\section{Electronic transitions of neutral C$_{60}$ in fullerene PNe}

The strongest allowed electronic transitions of neutral gas phase C$_{60}$
molecules, as measured in laboratory experiments, are located at 3760, 3980,
and 4024 \AA~with widths of 8, 6, and 4 \AA, respectively (Sassara et al.
2001; see also Garc\'{\i}a-Hern\'andez et al. 2012b). Garc\'{\i}a-Hern\'andez
\& D\'{\i}az-Luis (2013) found no evidence of these strong
neutral C$_{60}$ optical bands in absorption (or emission) in the fullerene PN
Tc 1. 

The S/N in the M 1-20 optical spectrum is too low to search for neutral
C$_{60}$ features in its spectrum (Garc\'{\i}a-Hern\'andez \& D\'{\i}az-Luis
2013), but here we have searched the higher S/N ($\sim$100 in the continuum
around 4000 \AA) spectrum of the PN IC 418 for the strongest electronic
transitions of neutral C$_{60}$ mentioned above. As in the case of Tc 1, we
can find no evidence of neutral C$_{60}$ in absorption (or
emission) around the expected wavelengths of $\sim$3760, 3980, and 4024 \AA.
This is shown in Figure 10 where we display the IC 418 velocity-corrected
spectra around the most intense C$_{60}$ transitions in comparison with those
of Tc 1 and its comparison star HR 6334. We note that several strong O lines
and He I 4026 \AA~very likely prevent identification of any broad and weak
absorption feature around 3760 and 4024 \AA, respectively. However, there is
no evidence of the neutral C$_{60}$ feature at 3980 \AA, a
wavelength region that is free of other spectral features. 

The one-sigma detection limits on the EQWs derived from our IC 418 spectrum
are 202, 75, and 44 m\AA\ for the 3760, 3980, and 4024 \AA~neutral C$_{60}$
transitions, respectively. By using the Spitzer formula (N$\sim$10$^{20}$
$\times$ (EQW/($\lambda$$^{2}$ $\times$ f))), where f is the oscillator
strength of each C$_{60}$ transition (Sassara et al. 2001), this translates
into column densities of $\sim$2 $\times$ 10$^{13}$, 4 $\times$ 10$^{13}$, and
2 $\times$ 10$^{13}$ cm$^{-2}$. These column density limits are similar to
those previously obtained in Tc 1 by Garc\'{\i}a-Hern\'andez \& D\'{\i}az-Luis
(2013). By following the latter work, we could in principle compare these
column-density limits with estimates of the circumstellar density of C$_{60}$
molecules as derived from the IR C$_{60}$ emission bands. 

Unfortunately, only two C$_{60}$ IR bands (those at $\sim$17.4 and 18.9
$\mu$m) were covered by the IC 418 {\it Spitzer} spectrum, making the estimation of the number of C$_{60}$ molecules (N(C$_{60}$)) and
excitation temperature (T(C$_{60}$)) from the Boltzmann excitation diagram rather uncertain.
Such column-density estimates are very sensitive to T(C$_{60}$), and this
temperature is very poorly constrained in IC 418 given that the short
wavelength C$_{60}$ bands are not available and that the C$_{60}$ 17.4 $\mu$m
band is blended with PAH emission. In addition, the derivation of the C$_{60}$
column density based on the IR-emission spectrum relies on the assumption that
C$_{60}$ is at thermal equilibrium, and this is still an open question (see,
e.g., Bernard-Salas et al. 2012).

\section{A fullerene - diffuse band connection?}

Most of the strongest and well-studied DIBs, as well as other weaker DIBs
towards Tc 1, are found to be normal for its reddening. This indicates that the
carriers of these ``normal'' DIBs are not particularly overabundant towards
fullerene PNe. The exceptions are the DIBs at 4428, 5780, 6203, 6284, and 8621
\AA, which are found to be unusually strong towards Tc 1. The radial velocities
of the 5780, 6203, 6284, and 8621 \AA~features confirm their insterstellar
origin, and the higher degree of ionization towards Tc 1 (in comparison with its
comparison star) suggests that their carriers may be ionized species. The 4428
\AA~feature, however, seems to be centred at the Tc 1's radial velocity,
suggesting a circumstellar origin (see Section 4). The situation is less clear
for M 1-20 and IC 418 because their comparison stars seem to map slightly
different ISM conditions (see Section 3.2), but at least the 4428 \AA~feature is
also found to be unusually strong in the latter fullerene PNe.  

The unusually strong 4428 \AA\ feature towards Tc 1 and M 1-20 prompted the
idea that the 4428 \AA\ carrier may be related to fullerenes or
fullerene-based molecules (Garc\'{\i}a-Hern\'andez \& D\'{\i}az-Luis 2013).
Our new finding of an unusually strong 4428 \AA\ feature towards IC 418
suggests that this may be a common characteristic of fullerene PNe, reinforcing
the speculation of a possible fullerene-DIB connection. 

Our detection of DCBs at 4428 and 5780 \AA~in an environment rich in
fullerenes and fullerene-related molecules would inevitably provide a link
between fullerene compounds and the DIB carriers. Photo-absorption theoretical
models of several large fullerenes (such as C$_{80}$, C$_{240}$, C$_{320}$,
and C$_{540}$) and multi-shell fullerenes (carbon onions like
C$_{60}$@C$_{240}$, C$_{60}$@C$_{240}$@C$_{540}$) predict their strongest
optical (4000$-$7000 \AA) transitions very close to 4428 and 5780
\AA~(Iglesias-Groth 2007), suggesting they are possible carriers. 

We can estimate abundances of the carriers of the DCBs at 4428 and 5780 \AA~in
PN Tc 1. For a Tc 1 carbon abundance of 4.7 $\times$ 10$^{-4}$ (relative to H;
e.g., Garc\'{\i}a-Hern\'andez et al. 2012a), the fraction of elemental carbon
(f$_{C}$) that is locked in the carrier molecule M can be expressed as (see,
e.g., Tielens 2005; Cami 2014)

\begin{equation} \label{bt1} f_{C}  \sim 9.93 \times 10^{-3} \times
\frac{W_{\lambda}}{E_{B-V}} \times \frac{N_{C}}{60} \times
\frac{5000^{2}}{\lambda^{2}} \times \frac{10^{-2}}{f}, \end{equation}

where W$_{\lambda}$/E$_{B-V}$ is the equivalent width per reddening unit (in
\AA~magnitude$^{-1}$), N$_{C}$ is the number of carbon atoms, and $\lambda$
and f are the transition wavelength (in \AA) and oscillator strength,
respectively. 

Assuming f=0.01 (Watson 1994; Weisman et al. 2003) and N$_{C}$$\geq$60 (as
appropiate for large fullerenes and buckyonions), we find the well known
result that the 4428 and 5780 \AA~carriers have to be very abundant or else they
have larger oscillator strengths (Tielens 2005). For example,
f$_{C}$$\sim$0.06, 0.19, and 0.42 for C$_{80}$, C$_{240}$, and C$_{540}$,
respectively, if these fullerene species are considered to be the only carrier
of the 4428 \AA~feature\footnote{Similar values are obtained for the
4428 \AA~feature in the other fullerene PNe M 1-20 and IC 418.}. In Tc 1, the
fraction of elemental carbon that is locked in C$_{60}$ is estimated to be
$\sim$4 $\times$ 10$^{-4}$ (as estimated from the IR emission; e.g.,
Garc\'{\i}a-Hern\'andez et al. 2012a). However, most fullerenes bigger than
C$_{60}$ and multi-shell fullerenes (buckyonions) display strong transitions
close to 4428 \AA~(Iglesias-Groth 2007). Thus, in the fullerene-DIB
hypothesis, the broad 4428 \AA~feature would be the result of the
superposition of the transitions of a series (family) of fullerenes bigger
than C$_{60}$ and buckyonions, and each fullerene compound would contribute to
the total EQW observed. Unfortunately, at present the possible relative
contribution (e.g., in terms of FWHM and EQW) of these fullerene compounds to
the 4428 \AA~feature is not known. 

More interesting is that only C$_{540}$ and
C$_{60}$@C$_{240}$@C$_{540}$ display a strong transition near 5780
\AA~(Iglesias-Groth 2007). By considering the latter fullerene species (and
using f=0.01) as the only carrier of the 5780 \AA~feature,
f$_{C}$$\sim$2$\times$10$^{-3}$ and 3$\times$10$^{-3}$ are obtained for
C$_{540}$ and C$_{60}$@C$_{240}$@C$_{540}$, respectively. Thus, a greater
oscillator strength (e.g., f=0.1) for the C$_{540}$ and
C$_{60}$@C$_{240}$@C$_{540}$ transitions at 5780 \AA~would decrease the latter
estimates to levels similar to C$_{60}$. On the other hand, it is to be noted
here that the C$_{60}$ abundance estimation in Tc 1 (e.g.,
Garc\'{\i}a-Hern\'andez et al. 2012a) assumes a uniform C$_{60}$ spatial
distribution in the circumstellar shell. Bernard-Salas et al. (2012) present
evidence that the C$_{60}$ emission in Tc 1 is extended and peaks far away from
the central star. If the C$_{60}$ molecules in Tc 1 are indeed distributed in a
ring (or in clumps) around the central star, then the quoted C$_{60}$ abundance
of $\sim$4 $\times$ 10$^{-4}$ (relative to C; e.g., Garc\'{\i}a-Hern\'andez et
al. 2012a) should be considered as a lower limit. This may increase the
C$_{60}$ abundance estimate (from the IR emission) in Tc 1 to values higher than
the 5780 \AA~carrier/s abundance estimated here. In addition, it may explain the
lack of the strongest optical electronic transitions of the C$_{60}$ molecule in
Tc 1 (and IC 418).

From the previous paragraghs, we conclude that at present large fullerenes and
buckyonions cannot be completely discarded as possible carriers of the 4428
~and 5780 \AA~features. However, another fullerene-related species should be
considered as possible diffuse band carriers (see below).   

Recent experimental studies demonstrate that fullerenes (and
metallofullerenes) would react with polycyclic carbon, graphene-like
structures, and PAHs, forming a rich family of fullerene-based molecules such
as fullerene/PAH clusters and endohedral metallofullerenes (Dunk et al. 2013).
These fullerene-related species may still be excited by the UV photons from
the central star, emitting through the same IR vibrational modes as empty
cages. Laboratory work shows that fullerene/PAH adducts (such as
C$_{60}$/anthracene) can easily form via Diels-Alder cyclo-addition
reactions, displaying mid-IR features strikingly coincident with those from
neutral C$_{60}$ and C$_{70}$ (Garc\'{\i}a-Hern\'andez, Cataldo \& Manchado
2013). In addition, gas-phase reactions between PAHs and C$_{60}$ and C$_{70}$
are experimentally proven to occur under circumstellar/interstellar conditions
(Dunk et al. 2013), and the resulting reaction products (e.g., C$_{70}$-PAH
cluster ions like  C$_{70}$C$_{24}$H$_{10}$$^{+}$) are very stable. Metals
such as Na (and Ca) are also quite abundant in the fullerene-rich
circumstellar envelope of Tc 1 and metallofullerene (e.g., Na@C$_{60}$)
formation is expected to be as efficient as empty fullerenes (Dunk et al.
2013). Indeed, theoretical spectra of Na@C$_{60}$ (Dunk et al. 2013) show the
same four vibrational modes as neutral C$_{60}$ (but with much higher
absorption intensities), together with a new IR-active vibrational mode due to
the metal encapsulation. Interestingly, the wavelength position of this new
mode is quite close to the still unidentified $\sim$6.4 $\mu$m feature
observed in Tc 1 and other fullerene-rich PNe (Dunk et al. 2013;
Garc\'{\i}a-Hern\'andez et al. 2010, 2011b, 2012a; Bernard-Salas et al.
2012). 

Certainly, metallofullerenes are better diffuse band carrier candidates than
fullerene/PAH adducts because the latter species are less stable towards UV
radiation (see, e.g., Kroto \& Jura 1992). In particular, adducts of C$_{60}$
with linearly condensed PAHs (acenes such as anthracene, tetracene, and
pentacene; e.g., Garc\'{\i}a-Hern\'andez, Cataldo \& Manchado 2013) are not as
stable as those with pericondensed PAHs (e.g., coronene); under the action of
strong UV radiation (e.g., from the central star), the C$_{60}$/acene adducts
may be dissociated back to the starting molecules. However, if the
C$_{60}$/acene adducts are formed in a circumstellar region shielded from the
UV radiation (or they are absorbed by dust particles), then they could survive
in PNe circumstellar shells.

In short, fullerenes in their multifarious manifestations - buckyonions,
fullerene clusters, fullerenes-PAHs, metallofullerenes, fullerene-like
fragments or buckybowl structures - may help solve the long-standing
astrophysical problem of the identification of some of the DIB carriers. Our
detection of DCBs at 5780 and 4428 \AA~in Tc 1 may thus help to identify the
carrier molecule(s), so more theoretical/laboratory work on fullerene-related
molecules is encouraged.

  \begin{figure*}
   \centering
   \includegraphics[angle=0,scale=.45]{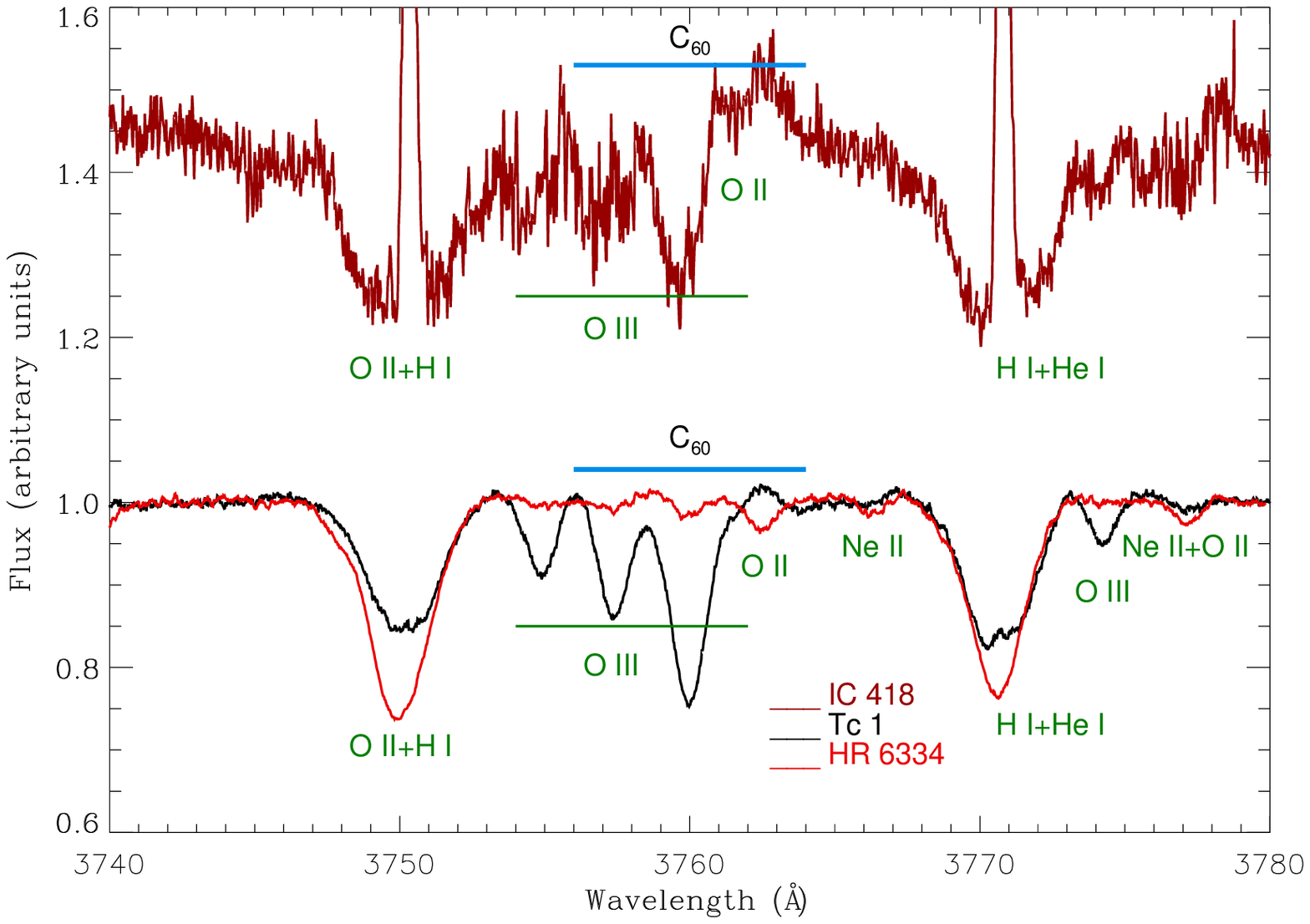}%
   \includegraphics[angle=0,scale=.45]{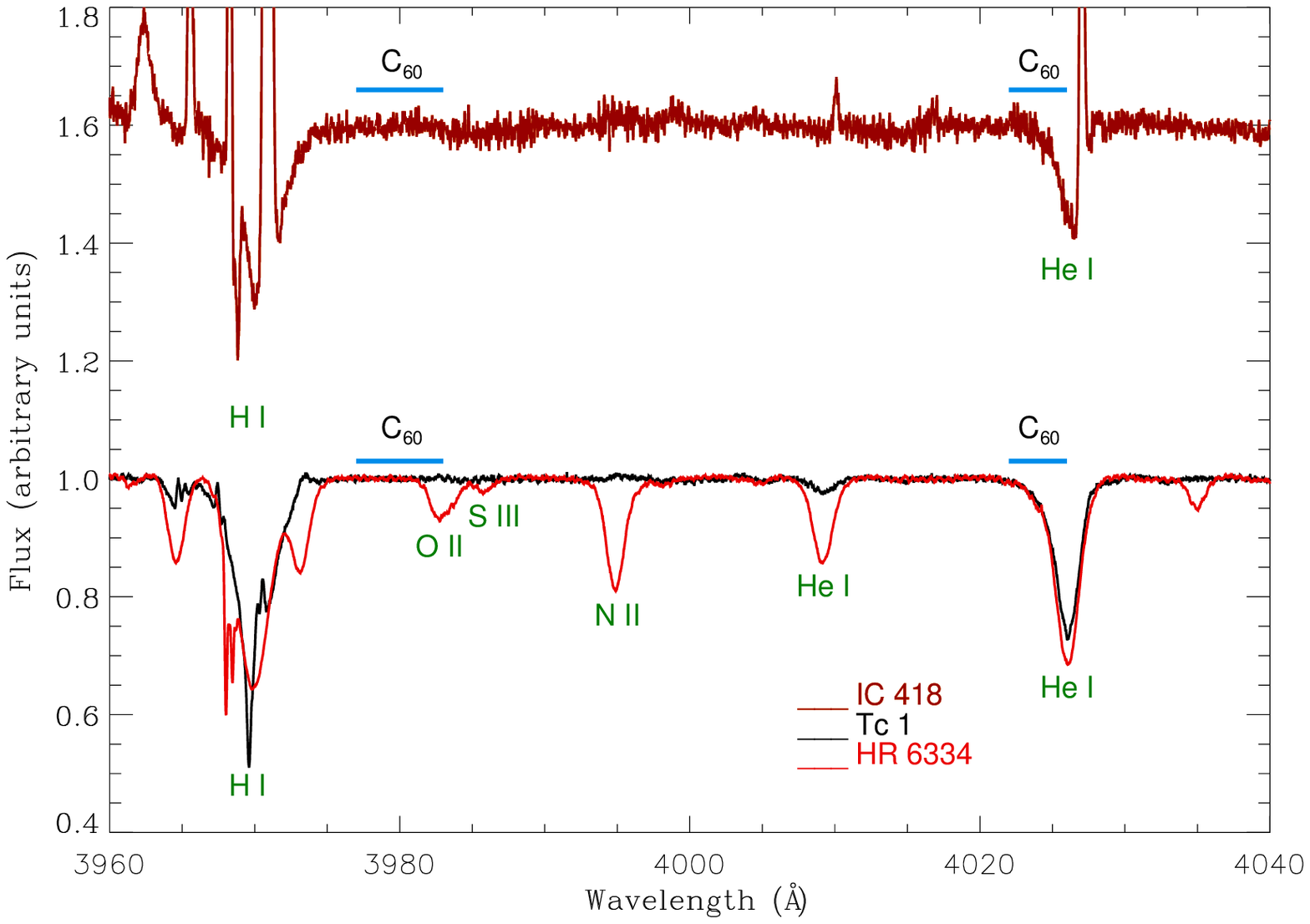}
    \caption{Velocity-corrected spectra of IC 418 (in brown), Tc 1 (in black)
    and HR 6334 (in red) around 3760 \AA\ (left panel) and 4000 \AA\ (right
    panel) where the atomic line identifications are indicated (in green). The
    expected positions (and FWHMs) of the C$_{60}$ features are indicated on top of
    the spectra. There is no evidence (additional absorption and/or emission) in
    IC 418 and Tc 1 for the neutral C$_{60}$ features at 3760, 3980, and 4024
    \AA.
\label{Fig10}}
    \end{figure*}

\section{Conclusions}

We have searched DIBs in the optical spectra towards three fullerene-containing PNe
(namely Tc 1, M 1-20, and IC 418). We have identified 20, 12, and 11 DIBs
towards Tc 1, M 1-20, and IC 418, respectively. All of these absorption bands
are known DIBs as previously reported in the literature.  

Towards Tc 1, the PN with the highest S/N spectrum and a proper comparison star,
six of the strongest and well-studied DIBs (i.e., those at 5797, 5850, 6196,
6270, 6379, and 6614 \AA), and nine other weaker interstellar features (i.e.,
those at 5776, 6250, 6376, 6597, 6661, 6792, 7828, 7833, and 8038 \AA) are found
to be normal for its reddening. This indicates that the carriers of these
``normal'' DIBs are not particularly overabundant towards fullerene PNe. The
five DIBs at 4428, 5780, 6203, 6284, and 8621 \AA~are found to be unusually
strong in the Tc 1 line of sight. The radial velocities of the 5780, 6203, 6284,
and 8621 \AA~features confirm their insterstellar origin, and the high
ionization degree towards Tc 1 suggests that their carriers may be ionized
species. The 4428 \AA~feature, however, seems to be centred at the Tc 1's radial
velocity, suggesting a circumstellar origin. 

The situation is less clear for the fullerenes PNe M 1-20 and IC 418, because
their spectra are of lower quality than in Tc 1, and their comparison stars
seem to map slightly different ISM conditions. In spite of this, the same
classification scheme (``normal'' versus ``unusually strong'' DIBs) seems to
be applicable towards M 1-20 and IC 418. At least the 4428 \AA~feature is
found to be unusually strong in these objects, as a common characteristic
to fullerene PNe.

The Tc 1's high radial velocity permitted us to search its high-quality
optical spectrum for DCBs. Interestingly, we report the first tentative
detection of two DCBs at 4428 and 5780 \AA~in the fullerene-rich circumstellar
environment around Tc 1. The presence of 4428 and 5780 \AA\ nebular emission
at the radial velocity of Tc 1 further suggests their circumstellar origin.
The non-detection of DCBs in the other fullerene PNe is due to the low S/N in
our optical spectra.

Moreover, we can find no evidence of the strongest electronic
transitions of neutral C$_{60}$ in the IC 418 optical spectrum. The
non-detection of neutral C$_{60}$ optical absorptions in fullerene PNe could
be explained if the C$_{60}$ IR emission peaks far away from the central star.
Mid-IR images at high spatial resolution and centred on the C$_{60}$
features would be desirable to understand the lack of the C$_{60}$ optical
bands in fullerene-containing PNe.

We have estimated the abundances of the carriers of the DCBs at 4428 and 5780
\AA, and we conclude that at present large fullerenes and buckyonions cannot be
completely discarded as possible carriers of the 4428 and 5780 \AA~features. 

On the basis of detecting DCBs at 4428 and 5780 \AA\ in Tc 1, we
suggest that laboratory and theoretical studies of fullerenes in their
multifarious manifestations - buckyonions, fullerene clusters,
fullerenes-PAHs, metallofullerenes, fullerene-like fragments or buckybowl
structures - may help solve the astronomical mistery of the identification of
some of the DIB carriers. 


\begin{acknowledgements}

We acknowledge the anonymous referee for very useful suggestions that helped
to improve the paper. We also acknowledge Jack Baldwin, Robert Williams, and
Mark Phillips for supplying us with the nebular spectrum of Tc 1, as well as
Jorge Garc\'{\i}a-Rojas for his help during the data analysis. N.K.R. thanks
the Instituto de Astrof\'{\i}sica de Canarias for inviting him as a Severo
Ochoa visitor during January to April 2014 when part of this work was done.
J.J.D.L., D.A.G.H., and A.M. acknowledge support provided by the Spanish
Ministry of Economy and Competitiveness (MINECO) under grant
AYA$-$2011$-$27754. D. A. G. H. also acknowledges support provided by the
MINECO grant AYA$-$2011$-$29060. This work is based on observations obtained
with ESO/VLT under the programme 087.D-0189(A). This article is also partially
based on service observations made with the Nordic Optical Telescope operated
on the island of La Palma by the Nordic Optical Telescope Scientific
Association in the Spanish Observatorio del Roque de Los Muchachos of the
Instituto de Astrof\'{\i}sica de Canarias. 

\end{acknowledgements}

\Online
\begin{appendix}
\section{Tables A1, A2, A3, A4, A5, and A6.}

 \begin{table*}
\tiny
\caption{\label{t2}Diffuse interstellar bands in Tc 1 and HR 6334.\tablefootmark{a}}
\centering
\begin{tabular}{lccccccccccc}
\hline\hline
 Tc 1          &     &         &     &     &            &HR 6334           &   &      &     &     &           \\
    $\lambda$$_{c}$            & Components   &    FWHM & EQW & S/N & EQW/E$_{B-V}$ & $\lambda$$_{c}$    & Components & FWHM & EQW & S/N & EQW/E$_{B-V}$ \\
($\AA$)         &   ($\AA$)    & ($\AA$) & (m$\AA$) & &  ($\AA$/mag) &  ($\AA$)   & ($\AA$)  & ($\AA$) & (m$\AA$)  &     &  ($\AA$/mag)  \\
\hline
4427.51\tablefootmark{b,c}  &  $\dots$  & 19.35 & 860.0 & 403    & 3.74   & 4429.71 & $\dots$  & 23.42 &  470.2   & 391    & 1.12  \\
5776.22  &  $\dots$   & 1.14 &  7.2  & 402   &0.03    & 5776.11  &  $\dots$   & 1.21 & 5.3   &    535    &  0.01      \\ 
5780.65\tablefootmark{c}  &  $\dots$      & 2.02  & 112.1  & 420   &0.49  & 5780.59 & $\dots$  & 2.17      &   119.0   &  586   & 0.28	  \\
5797.21 &  $\dots$       & 0.82 & 25.7 & 536  &0.11    & 5797.19 & $\dots$  & 1.15      &  45.7 & 590    & 0.11     \\
5849.74 &   $\dots$      & 0.89 & 2.2 & 437  &0.01    & 5849.74  & $\dots$ & 0.76	   & 12.7   &    536   & 0.03	   	\\
 6196 &  6196.06\tablefootmark{d}  &  1.11    & 10.7     &  525      & 0.05     &  6196 & 6196.05\tablefootmark{d} & 0.96     &  16.0  &   692  &   0.04       \\   
     & 6195.88 & 0.48 &  6.0 &    &0.03  &   &  6195.89  & 0.40 & 10.4   &      & 0.02     \\
     & 6196.54 & 0.39 &  2.6  &   &0.01 &   &  6196.56  & 0.36 & 4.3    &        & 0.01  	  \\
6203.33 & $\dots$	& 2.03 & 36.7 & 560  & 0.16    & 6203.17 & $\dots$  & 1.28   &  29.1   &  758      & 0.07	   \\
6250.36 & $\dots$      & 1.74 &  8.7 &  610 & 0.04    & 6250.62 & $\dots$  & 2.45    &  14.7   &  727      & 0.03	   \\
6270.08 & $\dots$    & 1.66 & 15.0 &  618  &0.06    & 6270 & 6269.98    & 1.47      &  25.8   & 494    & 0.06   \\
6284.18 &  $\dots$       & 4.41 & 630 &  380 & 1.66    & 6284.15 & $\dots$  & 4.55      &  309.8   &  693   & 0.74   \\
6376 & 6376.22\tablefootmark{d} &  1.38    &  2.6    &  559    &  0.01     &  6376 &  6376.40\tablefootmark{d} & 1.36 & 12.7 &  612 &    0.03 \\
      & 6376.05      & 0.57 &  1.7 &   &0.01   &   & 6375.92   & 0.41      &  3.6   &     & 0.01	  \\
     & 6376.76      & 0.48&  1.6 &   &0.01  &  & 6376.65   & 0.76    &  6.4   &    & 0.01	  \\
6379 & 6379.41\tablefootmark{d}    & 1.13  &  8.7    &  625    &  0.04     & 6379 & 6379.47\tablefootmark{d} &   1.26        &   56.4     &  812      &    0.13        \\
     & 6379.14      & 0.56 & 4.9 &   &0.02   &     & 6379.14   & 0.46     & 21.0   &     & 0.05     \\
     & 6379.82      & 0.93 & 4.7 &      &0.02  &  & 6379.78   & 1.08   & 35.4   &     & 0.08	   \\
6597.18 &  $\dots$     & 0.48  & 4.1 & 529  &0.02    & 6597.14 & $\dots$  & 0.45      &  5.0   &  338   & 0.01	   \\
6613.77 &  $\dots$     & 1.47  & 32.5 & 432  &0.14    & 6613.74 & $\dots$  & 1.40   & 91.8   &  464     & 0.22	  \\
6661 & 6661.09\tablefootmark{d}    & 1.12    &  3.5    &  341    &  0.01     & 6661 &  6660.81\tablefootmark{d} &  1.09 &  22.8      &  337      &   0.05         \\
     & 6660.64  & 0.64 &  2.5 &   &0.01  &    & 6660.61   & 0.54  & 14.4   &       & 0.03	      \\
     & 6661.40  & 0.24 &  1.9 &   &0.01  &   & 6661.35   & 0.30  &  5.4   &         & 0.01	       \\
6792.22  &  $\dots$    & 1.16  &  12.6 & 360  &0.05    & 6792.07 & $\dots$  & 1.10   & 15.2   & 329      & 0.04	   \\ 
7828.75  &  $\dots$    & 1.49 &   8.8 & 382  &0.04    & 7828.45  & $\dots$ & 1.07   & 6.1   &   572    & 0.01	  \\   
7832.50  &  $\dots$    & 1.76 &  10.3 & 475  &0.04    & 7832.63  & $\dots$ & 2.11   & 14.5   &  636   & 0.03	  \\   
8038.14  & $\dots$     & 0.96 &   5.4 &  396 &0.02    & 8038.01  & $\dots$ & 0.81   & 7.0   &  460     & 0.02	  \\   
8621.12  &  $\dots$    & 4.52 &  63.4 & 348  &0.28    & 8621.15  & $\dots$ & 4.50   & 51.7  &  626     & 0.12	 \\   
\hline
\end{tabular}
\tablefoot{
\\
\tablefoottext{a}{The 3-$\sigma$ errors in the EQWs scale as $\sim$3 $\times$
FWHM/(S/N), while we estimate that the FHWMs in Tc 1 are precise to the 0.03 \AA\
level (less for M 1-20 and IC 418).}
\\
\tablefoottext{b}{The parameters of this DIB are estimated by adopting a
Lorentzian profile (see, e.g., Snow et al. 2002).}
\\
\tablefoottext{c}{Circumstellar absorption features (or diffuse circumstellar
bands) may be possibly detected at these wavelengths (see Section 4).}
\\
\tablefoottext{d}{Undeblended DIB.}
}
\end{table*}

\begin{table*}
\tiny
\caption{\label{t3}Diffuse interstellar bands in M 1-20 and HR 6716.\tablefootmark{a}}
\centering
\begin{tabular}{lccccccccccc}
\hline\hline
M 1-20          &             &     &     &            &HR 6716           &   &         &     &           \\
  $\lambda$$_{c}$            &    FWHM & EQW & S/N & EQW/E$_{B-V}$ & $\lambda$$_{c}$  & FWHM & EQW & S/N & EQW/E$_{B-V}$ \\
($\AA$)             & ($\AA$) & (m$\AA$) & &  ($\AA$/mag) &  ($\AA$)   & ($\AA$) & (m$\AA$)  &     &  ($\AA$/mag)  \\
\hline
4426.56\tablefootmark{b}    & 19.94\tablefootmark{c} & 2579.0\tablefootmark{c} & 20\tablefootmark{d}    & 3.22   & 4429.27   & 22.25 &  595.8   & 233    & 2.71  \\
5780.44    & 1.93  & 361.1  & 36   &   0.45  & 5780.41  & 2.01      &   164.5   &  792   & 	0.75  \\
5796.97    &  0.78    &  153.4     & 39  &  0.19          & 5796.95   & 0.74      & 37.2   &  661      & 0.17 \\
5849.69    & 0.95 &  70.7 &  52    &0.09  & 5849.69      & 0.84    & 9.3   &    639     &   0.04   \\
6195.83    &  0.43   &  44.3   & 71  &   0.05   & 6195.82   & 0.41    &  13.2   &  612       &   0.06     \\
6203.00    & 1.39 &  97.7 &  77     & 0.12    & 6203.02     & 1.25   &  28.9   & 688       & 	0.13   \\
6269.89    & 1.51 &  89.2 &  67     & 0.11    & 6269.70  & 1.02    &  12.4   &   684    & 0.06	   \\
6283.60    & 4.06 & 573.0 & 73      &0.72   & 6283.51   & 4.29      &  474.4   &  641     & 2.16  \\
6375.94    & 0.51 &  26.3 & 75    &0.03   & 6375.89    &  0.65    &  6.3       &  656       &   0.03         \\
6379.14    & 0.59   & 80.6    & 100  &   0.10        & 6379.12   &  0.63     & 22.5    &   687     &  0.10       \\
6613.48    & 1.01  & 177.0 &    79   & 0.22    & 6613.41  & 0.86   &  37.5   &  581     &   0.17 	    \\ 
6660.52    & 0.59  & 27.5 &  74     &0.03    & 6660.55   & 0.41      &  5.0   &  531     & 0.02	   \\	   
\hline
\end{tabular}
\tablefoot{
\\
\tablefoottext{a}{The 3-$\sigma$ errors in the EQWs scale as $\sim$3 $\times$
FWHM/(S/N), while the FHWMs are less precise than 0.03 \AA.}
\\
\tablefoottext{b}{The parameters of this DIB are estimated by adopting a
Lorentzian profile (see, e.g., Snow et al. 2002). The central
wavelength in M 1-20 is very uncertain.}
\\
\tablefoottext{c}{Best estimates found by clipping out the narrow emission lines
and smoothing the spectrum with boxcar 15. The error in the quoted EQW is
estimated to be $\sim$786 m\AA.}
\\
\tablefoottext{d}{S/N in the original spectrum.}
}
\end{table*}

\begin{table*}
\tiny
\caption{\label{t4}Diffuse interstellar bands in IC 418 and HR 1890.\tablefootmark{a}}
\centering
\begin{tabular}{lccccccccccc}
\hline\hline
IC 418           &      &     &     &            &HR 1890           &      &     &     &            &   Hobbs et al. (2008)  &   Luna et al.(2008) \\
$\lambda$$_{c}$ & FWHM & EQW & S/N & EQW/E$_{B-V}$ &   $\lambda$$_{c}$           & FWHM & EQW & S/N & EQW/E$_{B-V}$ & EQW/E$_{B-V}$ & EQW/E$_{B-V}$\\
($\AA$)         &($\AA$) & (m$\AA$) & &  ($\AA$/mag)       &($\AA$) & ($\AA$) & (m$\AA$)  &     &  ($\AA$/mag)          &   ($\AA$/mag &  ($\AA$/mag)    \\
\hline
4426.20\tablefootmark{b} & 21.45     & 1001.0    &  116      &4.35    & $\dots$  & $\dots$ &  $\dots$         &  $\dots$   &        $\dots$ & 1.10   &  $\dots$  \\
5780.94 & 2.02  & 99.8 &195 &0.43    & 5780.97   & 1.68      &  25.5       &344       & 0.32      &  0.23    &  0.46     \\
5797.51 & 0.85  & 34.8 &184 &0.15    & 5797.46   & 1.07      &  8.4       &336       & 0.10      &  0.18    &  0.17     \\
5850.55 & 1.15  & 30.9 &175 &0.13    & 5850.44   & 1.03      &  7.1       &351       & 0.09      &  0.09   &  0.061    \\
6196.47 & 0.31  & 10.0 &166 &0.04    & 6196.27  & 0.29      &  2.9       &   343       & 0.04      &    0.03     &   0.053          \\
6203.37 & 0.80  & 15.3 &171 &0.07    & $\dots$   & $\dots$   &  $\dots$  &  $\dots$ & $\dots$   &  0.05   &  $\dots$    \\
6270.24 & 1.08  & 15.7  & 183 &0.07  & $\dots$	& $\dots$   & $\dots$	& $\dots$  & $\dots$	& 0.07   & $\dots$      \\
6284.30 & 4.88  & 218.8 & 183 &0.95  & $\dots$   & $\dots$   &  $\dots$  &  $\dots$ & $\dots$   &  0.41    &  0.90     \\
6376.41 & 0.35  &  2.9  & 137 &0.01  & $\dots$	& $\dots$   & $\dots$	& $\dots$  & $\dots$	& 0.04  & $\dots$      \\
6379.94 & 0.43  &  5.2  & 132 &0.02  & $\dots$	& $\dots$   & $\dots$	& $\dots$  & $\dots$	& 0.08  & 0.088   \\
6614.11 & 1.04  & 30.8  & 146 &0.13  & 6614.20   & 0.91      &   6.6       & 364      & 0.08      &  0.15    &  0.21     \\
\hline
\end{tabular}
\tablefoot{
\\
\tablefoottext{a}{The 3-$\sigma$ errors in the EQWs scale as $\sim$3 $\times$
FWHM/(S/N), while the FHWMs are less precise than 0.03 \AA.}
\\
\tablefoottext{b}{The parameters of this DIB are estimated by adopting a
Lorentzian profile (see, e.g., Snow et al. 2002). The quoted central
wavelength in IC 418 is quite uncertain.}
}
\end{table*}

\begin{table*}
\caption{Radial velocities (in kms$^{-1}$) for the atomic and molecular lines as
well as DIB features with two interstellar components in Tc 1 and HR 6334.\tablefootmark{a}}
\centering
\small\begin{tabular}{lcccccc}
\hline\hline
\underline{Feature}  &  & \underline{Tc 1 } &  &  &  \underline{HR 6334} &  \\
                     &  CS &  & IS & & IS & \\
\hline
Na I         & $-$116.10 & $-$83.00 & $-$4.67 & 26.36 & $-$4.67 & 26.36 \\
Ca I         &          &         & $-$7.16 & 22.98 &         &       \\
CH$^{+}$     &          &         & $-$6.87 & 25.50 & $-$2.83 & 31.24 \\
CH           &          &         &         &       & $-$5.86 & 27.96 \\
CN (R1)      &          &         &         &       & $-$5.34 & (42)  \\
CN (R0)      &          &         &         &       & $-$4.57 & 29.25  \\
CN (P1)      &          &         &         &       & $-$4.18 & 31.02 \\
DIB 6196\AA\ &          &         &$-$8.03  & 25.11 & $-$5.42 & 27.68 \\
DIB 6376\AA\ &          &         &$-$3.86  & 30.00 &$-$7.05  & 33.34 \\
DIB 6379\AA\ &          &         &$-$6.58  & 23.97 &$-$5.78  & 29.18 \\
\hline\hline
\end{tabular}
\tablefoot{
\\
\tablefoottext{a}{Typical uncertainty of $\sim$ $\pm$ 1 kms$^{-1}$.}
}
\end{table*}

\begin{table*}
\caption{Radial velocities (in kms$^{-1}$) for the atomic lines as
well as DIB features in M 1-20 and HR 6716.\tablefootmark{a}}
\centering
\small\begin{tabular}{lcccccccccc}
\hline\hline
\underline{Feature} &  & \underline{M 1-20} &  \underline{HR 6716} & \\
        &   CS             & IS  & IS &    \\
\hline
Na I  &    60.95?  & $-$6.54 &  $-$26.43   & $-$6.12  \\
DIB 5797\AA\    &     &  $-$5.19      &     &   $-$5.33    \\
DIB 6196\AA\    &     &   $-$7.85      &     &   $-$7.88   \\
DIB 6379\AA\    &     &   $-$8.63       &     &  $-$9.04   \\
\hline\hline
\end{tabular}
\tablefoot{
\\
\tablefoottext{a}{Typical uncertainty of $\sim$ $\pm$ 1 kms$^{-1}$.}
}
\end{table*}

\begin{table*}
\caption{Radial velocities (in kms$^{-1}$) for the atomic and molecular lines, as
well as DIB features in IC 418 and HR 1890.\tablefootmark{a}}
\centering
\small\begin{tabular}{lcccccc}
\hline\hline
\underline{Feature}  &  & \underline{IC 418}    &  \underline{HR 1890} &  \\
                     &  CS &   IS  & IS & \\
\hline
Na I         &  58.36   &           22.22         & 5.42    &  23.75 \\
CH           &          &           22.67        &         &        \\
DIB 5780\AA\ &          &           25.34         &         &  23.33 \\
DIB 5797\AA\ &          &           22.16         &         &  22.15 \\
DIB 5850\AA\ &          &           25.60        &         &  30.97 \\
\hline\hline
\end{tabular}
\tablefoot{
\\
\tablefoottext{a}{Typical uncertainty of $\sim$ $\pm$ 1 kms$^{-1}$.}
}
\end{table*}

\end{appendix}

\end{document}